\documentclass[%
 reprint,
 amsmath,amssymb,
 aps,
floatfix,
]{revtex4-2}

\usepackage{graphicx}
\usepackage{dcolumn}
\usepackage{bm}
\usepackage{hyperref}

 \usepackage{booktabs}
 \usepackage{xcolor}
\usepackage{graphicx}
\usepackage{slashed}

\begin{document}


\title{Investigating triply heavy tetraquark states through QCD sum rules}

\author{Wen-Shuai Zhang$^{\,1}$}
\author{Liang Tang$^{\,1}$}%
 \email{tangl@hebtu.edu.cn}
\affiliation{%
\textit{College of Physics and Hebei Key Laboratory of Photophysics Research and Application, }
\\
\textit{Hebei Normal University, Shijiazhuang 050024, China} 
}%





\begin{abstract}
We apply the method of QCD sum rules to study the  \(QQ\bar{Q}\bar{q}\) and \(QQ\bar{Q}\bar{s}\) tetraquark states, where $Q=c,b$ and $q=u,d$, with the  quantum number \(J^P = 0^{+}\). We consider the contributions of vacuum condensates up to dimension-9 in the operator product expansion, and use the energy scale formula \(\mu = \sqrt{M_{X}^2 - (i\mathbb{M}_c + j\mathbb{M}_b)^2} - k\mathbb{M}_s\) to determine the optimal energy scales for the QCD spectral densities. Our results indicate that triply charm tetraquark states \(cc\bar{c}\bar{q}\) and \(cc\bar{c}\bar{s}\) have masses in the ranges of $5.38-5.84\,\text{GeV}$ and $5.66-6.16\,\text{GeV}$, respectively. In the bottom sector, triply bottom tetraquark states \(bb\bar{b}\bar{q}\) and \(bb\bar{b}\bar{s}\) have masses in the ranges of $14.89-15.55\,\text{GeV}$ and $14.95-15.66\,\text{GeV}$, respectively. This study could help distinguish these states in upcoming high-energy nuclear and particle experiments.
\begin{description}
\item[Key\ words]
 exotic \ hadron, \ tetraquark \ state, \  QCD \ sum \ rules 
\end{description}
\end{abstract}

\maketitle


\section{\label{sec:level1}INTRODUCTION\protect}
In 1964, M. Gell-Mann\cite{ref1} and G. Zweig\cite{ref2} independently developed the quark model for hadron classification, consisting of $q\bar{q}$ or $qqq$, and with it, the concept of multiquark states was introduced. In quantum field theory, quantum chromodynamics (QCD) permits the existence of exotic hadronic states, such as multiquark states, hybrid states, and glueballs. Since the discovery of $X(3872)$\cite{ref3} in 2003, dozens of multiquark states or their candidates have been observed experimentally. However, the internal structures of many new hadronic states have not yet been conclusively determined\cite{ref118,ref119,ref120}. This makes the study of multiquark states not only important in terms of experimental research but also significant theoretically, both in the present and for the future. 

Recent observations of new hadronic states suggest that some are good candidates for QCD exotic states, which can be classified based on various criteria, such as quark composition, electric charge, interactions, and quantum numbers.\, Interestingly, \,most of these new states contain at least one heavy quark or antiquark. Due to the larger mass of heavy quarks, their properties are often discussed in the heavy quark limit. Based on the number of heavy quark components, these states can be classified into three types: those with one, two, or four heavy quarks/antiquarks. Next, we will review the representative references for each type.

\begin{itemize}
\item{ Systems that contain a single heavy quark or antiquark as a constituent.} The charmed-strange states \(D_{s0}(2317)\) and \(D_{s1}(2460)\), first observed by BaBar in 2003 \cite{ref4} and later confirmed by other experiments \cite{ref5, ref6}, have masses much lower than those predicted by traditional quark model \cite{ref10}, sparking interest in their internal structures \cite{ref7,ref8,ref9}. Interpretations include molecular states (\(DK\) or \(D^*K\)) \cite{ref11,ref12,ref13,ref14,ref15,ref16,ref17,ref18,ref19} or tetraquark states \cite{ref20,ref21,ref22,ref23,ref24,ref25, ref113}.

\item{Systems that contain two heavy quarks or antiquarks as constituents.} States containing two heavy quarks or antiquarks have attracted significant attention in recent decades. The \(X(3872)\), discovered by Belle in 2003 \cite{ref26,ref27,ref28,ref29}, exhibits quantum numbers consistent with charmonium \cite{ref30,ref31,ref32,ref33,ref34,ref35,ref36,ref37}, but its mass deviates from quark model predictions \cite{ref38}, and its proximity to the \(D\bar{D}^*\) threshold complicates interpretation. Suggested explanations include a molecular state \cite{ref39,ref40,ref41,ref42,ref43,ref44,ref45,ref46,ref47,ref48,ref49,ref50} or a tetraquark \cite{ref51,ref52,ref53,ref54,ref55,ref56,ref57,ref58,ref59,ref60,ref61,ref62,ref63,ref64,ref114}. Discoveries of \(Z_c(3900)\), \(Z_c(4025)\) \cite{ref116}, \(Z_b(10610)\) \cite{ref115}, and the pentaquarks \(P_c(4380)\) and \(P_c(4450)\) have further driven research into hidden-charm/bottom multiquark states.

\item{Systems containing four heavy quarks or antiquarks.} The X(6900), observed by LHCb in 2020, is the first confirmed all-heavy tetraquark state\cite{ref65}, consisting of two charm and two anti-charm quarks, with a mass around 6900 MeV. Similar states, X(6600) and X(7200), have been observed by CMS \cite{ref66} and ATLAS \cite{ref67} in the double \(J/\psi\) spectrum. These recent experimental advancements have sparked extensive theoretical investigations. The interpretations of their nature have been widely discussed in scenarios such as tetraquarks \cite{ref68,ref69,ref70,ref71,ref72,ref73,ref74,ref75,ref77} and gluonic tetraquarks \cite{ref76,ref123}, enhancing our understanding of strong interactions among heavy quarks and non-perturbative QCD.
\end{itemize}

The study of the existence of triply heavy tetraquark states \(QQ\bar{Q}\bar{q}\) is a natural extension of these theoretical investigations.
Recent studies \cite{ref78,ref79,ref80,ref81,ref82,ref83,ref84,ref85,ref86,ref87,ref88,ref117} have predicted the masses and other properties of triply heavy tetraquark states, although some controversies remain. Ref. \cite{ref78} investigated triply heavy tetraquark systems using the constituent quark model and $\text{GEM}+\text{CSM}$ method. Narrow resonances were found in the ranges of $5.6-5.9 \, \text{GeV}$ and $15.3-15.7 \, \text{GeV}$ for triply charm tetraquarks and triply bottom tetraquarks, respectively. Ref. \cite{ref85} studied triply heavy tetraquark states within the framework of QCD sum rules up to dimension five in the operator product expansion. The masses of the tetraquark states \(cc\bar{c}\bar{q}\) and \(bb\bar{b}\bar{q}\) with quantum number \(0^+\) were found to be \(5.1 \pm 0.2 \, \text{GeV}\) and \(13.5 \pm 0.4 \, \text{GeV}\), respectively. Within the error margins, the masses of the tetraquark states \(cc\bar{c}\bar{q}\) and \(bb\bar{b}\bar{q}\) with quantum number \(1^+\) were found to be the same as those with quantum number \(0^+\).

In this work, we apply $\text{QCD}$ sum rules to study triply heavy tetraquark states, using the energy scale formula \(\mu = \sqrt{M_X^2 - (iM_c + jM_b)^2 - kM_s}\) to determine the optimal $\text{QCD}$ spectral density scale and consider the contributions of vacuum condensates up to dimension$\text{-9}$ in the operator product expansion. This study offers a new mass spectrum that can provide valuable insights into the properties of triply heavy tetraquarks.

The paper is organized as follows: Sec.\,\hyperref[sec:Interpolating Currents]{II} defines the interpolating currents for \(J^P = 0^{+}\) triple heavy tetraquark states. Sec.\,\hyperref[sec:QCD_SUM_RULE_ANALYSIS]{III} presents the formalism of QCD sum rules, while Sec.\,\hyperref[sec:NUMERICAL_ANALYSES]{IV} provides the numerical analysis. We conclude with findings and implications in Sec.\,\hyperref[sec:CONCLUSIONS]{V}.

\section{Interpolating Currents}

\label{sec:Interpolating Currents} 
To study the triply heavy tetraquark states, we shall first construct the diquark-antidiquark type  interpolating currents. There are five independent diquark fields: $q_a^T C q_b$, $q_a^T C\gamma_5 q_b$, $q_a^T C\gamma_\mu q_b$, $q_a^T C\gamma_\mu \gamma_5 q_b$, and $q_a^T C\sigma_{\mu\nu} q_b$, where $a$ and $b$ denote color indices. In general, all of these diquarks and their corresponding antidiquarks can be used to construct tetraquark operators. In this paper, we use the diquark fields $q_a^T C\gamma_5 q_b$, $q_a^T C\gamma_\mu q_b$, $q_a^T C q_b$, and $q_a^T C\gamma_\mu \gamma_5 q_b$ to construct the tetraquark currents. By considering the Lorentz and color structures, we ultimately obtain interpolating currents with $J^P = 0^+$:
\begin{align}
J_1 &= \left(Q^\mathrm{T} _{1a}C\gamma_5 Q_{2b}\right)\left(\bar{Q} _{3a} \gamma_5 C \bar{q}^\mathrm{T}_{b} + \bar{Q} _{3b} \gamma_5 C \bar{q}^\mathrm{T}_{a}\right), \notag \\
J_2 &= \left(Q^\mathrm{T} _{1a}C\gamma_5 Q_{2b}\right)\left(\bar{Q} _{3a} \gamma_5 C \bar{q}^\mathrm{T}_{b} - \bar{Q} _{3b} \gamma_5 C \bar{q}^\mathrm{T}_{a}\right),
\notag \\
J_3 &= \left(Q^\mathrm{T} _{1a}C\gamma_\mu Q_{2b}\right)\left(\bar{Q} _{3a} \gamma^\mu C \bar{q}^\mathrm{T}_{b} + \bar{Q} _{3b} \gamma^\mu C \bar{q}^\mathrm{T}_{a}\right), \notag \\
J_4 &= \left(Q^\mathrm{T} _{1a}C\gamma_\mu Q_{2b}\right)\left(\bar{Q} _{3a} \gamma^\mu C \bar{q}^\mathrm{T}_{b} - \bar{Q} _{3b} \gamma^\mu C \bar{q}^\mathrm{T}_{a}\right), \notag \\
J_5 &= \left(Q^\mathrm{T} _{1a}C \gamma_\mu \gamma_5 Q_{2b}\right)\left(\bar{Q} _{3a}  \gamma_5 \gamma^\mu  C \bar{q}^\mathrm{T}_{b} + \bar{Q} _{3b}  \gamma_5 \gamma^\mu  C \bar{q}^\mathrm{T}_{a}\right), \notag \\
J_6 &= \left(Q^\mathrm{T} _{1a}C \gamma_\mu \gamma_5 Q_{2b}\right)\left(\bar{Q} _{3a}  \gamma_5 \gamma^\mu  C \bar{q}^\mathrm{T}_{b} - \bar{Q} _{3b}  \gamma_5 \gamma^\mu  C \bar{q}^\mathrm{T}_{a}\right), \notag \\
J_7 &= \left(Q^\mathrm{T} _{1a}C Q_{2b}\right)\left(\bar{Q} _{3a}  C \bar{q}^\mathrm{T}_{b} + \bar{Q} _{3b}  C \bar{q}^\mathrm{T}_{a}\right), \notag \\
J_8 &= \left(Q^\mathrm{T} _{1a}C Q_{2b}\right)\left(\bar{Q} _{3a}  C \bar{q}^\mathrm{T}_{b} - \bar{Q} _{3b}  C \bar{q}^\mathrm{T}_{a}\right), 
\label{Eq.(1)}
\end{align}
where the symbol \(q\) denotes the \(u/d/s\) quark field, and the symbol \(Q\) represents the \(c\) or \(b\) quark field.

The currents \(J_1\), \(J_3\), \(J_5\), and \(J_7\) exhibit color-symmetric structures \([6_c]_{Q_1Q_2} \otimes [\bar{6}_c]_{\bar{Q}_3\bar{q}}\), whereas \(J_2\), \(J_4\), \(J_6\), and \(J_8\) are characterized by color-antisymmetric structures \([\bar{3}_c]_{Q_1Q_2} \otimes [3_c]_{\bar{Q}_3\bar{q}}\). It should be noted that in the \(QQ\bar{Q}\bar{q}\) system, when the two heavy quarks in the diquark \(QQ\) are of identical flavor, such as \(cc\) or \(bb\), the currents \(J_2\), \(J_3\), \(J_6\), and \(J_8\) vanish due to Fermi-Dirac statistics \cite{ref110, ref111}.

In Sec.\,\hyperref[sec:QCD_SUM_RULE_ANALYSIS]{III}, we will proceed to incorporate these hadronic currents into the correlation function for calculation.

\section{QCD SUM RULE ANALYSIS}
\label{sec:QCD_SUM_RULE_ANALYSIS} 
In this paper, we study the triply heavy tetraquark states using the QCD sum rules. We investigate the correlation function associated with the scalar current:

\begin{equation}
\Pi(q) = i \int d^4x e^{iq \cdot x} \langle 0 | T[j(x) j^\dagger(0)] | 0 \rangle .
\label{Eq.(2)}
\end{equation}

At the hadronic level, the correlation function is represented through the spectral function via dispersion relation:

\begin{equation}
\Pi(q^2) = \int_{s_<}^{\infty} \frac{\rho_{\text{phen}}(s)}{s - q^2 - i \epsilon} ds.
\label{Eq.(3)}
\end{equation}

In this analysis, $s_< = 9m_Q^2$ is identified as the physical threshold. The spectral density $\rho_{\text{phen}}(s) \equiv \text{Im}\Pi(s)/\pi$, is parameterized to include the dominant pole contribution from the potential ground state $X \equiv |X; 0^{+}\rangle$, along with contributions from the continuum spectrum:

\begin{equation}
\begin{aligned}
    \rho_{\text{phen}}(s) & \equiv \sum_n \delta(s - M_n^2)\langle 0|J_{0^{+}}|n\rangle\langle n|J_{0^{+}}^\dagger|0\rangle \\
   & = \lambda_X^2 \delta(s - M_X^2) + \text{continuum}.
\label{Eq.(4)}
\end{aligned}
\end{equation}

Here, $M_X$ denotes the mass of the resonance X in its ground state, and $\lambda_X$ represents the residue at the pole.

Before discussing the quark-gluon level, let us introduce the Borel transformation:
\begin{equation}
\hat{B} \left[ f(q^2) \right] = \lim_{\substack{-q^2, n \to \infty \\ -q^2 / n \equiv M_B^2}}
 \frac{1}{n!} \left(-q^2\right)^{n+1} \left( \frac{d}{dq^2} \right)^n f(q^2).
\label{Eq.(5)}
\end{equation}

At the quark-gluon level, Eq. (\hyperref[Eq.(2)]{2}) can be computed using the operator product expansion (OPE), from which we can extract the spectral density $\rho_{\text{OPE}}(s)$, and for detailed spectral density, see the Appendix \hyperref[Appendix_A:_Spectral_densities_]{A}. To calculate Eq. (\hyperref[Eq.(2)]{2}) and obtain the Wilson coefficients, we will use the propagators of the $u/d$ quarks in coordinate space \cite{ref112}:
\begin{equation}
\begin{aligned}
S_{ab}^q(x) &= \frac{i \delta_{ab} \slashed{x}}{2\pi^2 x^4} - \frac{i t_{ab}^N G^{N}_{\alpha \beta}}{32 \pi^2 x^2} \left( \sigma_{\alpha \beta} \slashed{x} + \slashed{x} \sigma_{\alpha \beta} \right) \\
&- \frac{\delta_{ab}}{12} \langle \bar{q}q \rangle - \frac{\delta_{ab} x^2}{192} \langle \bar{q}g_s \sigma G q \rangle \\
&- \frac{t_{ab}^N \sigma_{\alpha \beta}}{192} \langle \bar{q}g_s \sigma G q \rangle. 
\label{Eq.(6)}
\end{aligned}
\end{equation}

The propagator for the $s$-quark in coordinate space is:
\begin{equation}
\begin{aligned}
S_{ab}^s(x) &= \frac{i \delta_{ab} \slashed{x}}{2\pi^2 x^4}-\frac{\delta_{ab} m_s}{4\pi^2 x^2} - \frac{\delta_{ab} x^2}{192} \langle \bar{s}g_s \sigma G s \rangle   \\
&- \frac{i t_{ab}^N G^{N}_{\alpha \beta}}{32 \pi^2 x^2} \left( \sigma_{\alpha \beta} \slashed{x} + \slashed{x} \sigma_{\alpha \beta} \right)+\frac{i\delta_{ab}\slashed{x}m_s}{48}\langle \bar{q}q \rangle\\
&- \frac{\delta_{ab}}{12} \langle \bar{q}q \rangle  +\frac{i\delta_{ab} x^2 \slashed{x}m_s}{1152} \langle \bar{s}g_s \sigma G s \rangle\\
&  +\frac{i t_{ab}^N m_s}{768}\left( \sigma_{\alpha \beta} \slashed{x} + \slashed{x} \sigma_{\alpha \beta} \right)\langle \bar{s}g_s \sigma G s \rangle \\
&- \frac{t_{ab}^N \sigma_{\alpha \beta}}{192} \langle \bar{s}g_s \sigma G s \rangle .
\label{Eq.(7)}
\end{aligned}
\end{equation}

In the above expression, the last two terms contain redundant $\alpha$ and $\beta$ indices, which are introduced to absorb gluons emitted by the heavy quark propagator, thereby forming quark-gluon condensates. Therefore, our approach will treat these two terms separately, and the detailed steps of the derivation are omitted.

The propagator for the heavy quark in coordinate space is:
\begin{equation}
\begin{aligned}
S_{ab}^Q(x) &=  i \int \frac{d^4 p}{(2\pi)^4} e^{-ip\cdot x}  \left\{ \frac{\delta_{ab}(\slashed{p}+m_Q)}{p^2-m_Q^2} \right. \\
& \left. -  g_s t^N_{ab} G^N_{\alpha \beta}  \frac{(\slashed{p}+m_Q)\sigma^{\alpha \beta}+\sigma^{\alpha \beta}(\slashed{p}+m_Q)}{4(p^2-m^2_Q)^2}  \right. \\ 
& \left.  + \frac{ m_Q ^2  (\slashed{p}+m_Q) \langle g_s^2 GG \rangle \delta_{ab}}{12(p^2-m^2_Q)^4}  \right. \\ 
& \left. + \frac{ m_Q \langle g_s^2 GG \rangle \delta_{ab}}{12(p^2-m^2_Q)^3}  \right\}.
\label{Eq.(8)}
\end{aligned}
\end{equation}

At the quark-gluon level, the correlation function can be expressed as follows:
\begin{equation}
\Pi(q^2) = \int_{s_<}^{\infty} \frac{\rho_{\text{OPE}}(s)}{s - q^2 - i \epsilon} ds,
\label{Eq.(9)}
\end{equation}
where:
\begin{equation}
\rho_{\text{OPE}}(s) \equiv \text{Im}\Pi(s)/\pi.
\label{Eq.(10)}
\end{equation}

Applying the Borel transformation to both sides of Eq. (\hyperref[Eq.(9)]{9}) yields:

\begin{equation}
\Pi(s_0, M_B^2)=\int_{s_<}^{\infty} e^{-s/M_B^2} \rho_{\text{OPE}}(s)\, ds,
\label{Eq.(11)}
\end{equation}
where:\, $\Pi(s_0, M_B^2)\equiv \hat{B}\left[ \Pi(q^2) \right]$.

We also apply the Borel transformation to the hadronic level and, by employing the principle of quark-hadron duality, derive the sum rule equation:

\begin{equation}
\lambda_X^2 e^{-M_X^2/M_B^2} = \int_{s_<}^{s_0} e^{-s/M_B^2} \rho_{\text{OPE}}(s)\, ds .
\label{Eq.(12)}
\end{equation}

This allows us to further derive:

\begin{equation}
M_{X}^2 = 
-\frac{\int_{9m_Q^2}^{s_0} ds \frac{\partial}{\partial\tau} \rho_{\text{OPE}}(s) \exp(-\tau s)}
{\int_{9m_Q^2}^{s_0} ds \rho_{\text{OPE}}(s) \exp(-\tau s)} \bigg|_{\tau = \frac{1}{M_{B}^2}}.
\label{Eq.(13)}
\end{equation}

For the choice of parameters $s_0$ and $M_B^2$, we will provide a detailed discussion in the following section.

\section{NUMERICAL ANALYSES}
\label{sec:NUMERICAL_ANALYSES}

In this paper, we neglect the masses of the light quarks $u$ and $d$. For the numerical analysis, we consider the running effects of the energy scale $\mu$, where the $\overline{MS}$ masses of the heavy quarks $m_c(\mu)$, the vacuum condensates, the quark-gluon spectral density $\rho(s, \mu)$, and the physical threshold  $9m_c^2(\mu)$ are all dependent on the energy scale $\mu$. At\,\,$\mu = 1 \,\text{GeV}$, the standard values for the vacuum condensates are given by:
$\langle \bar{q}q \rangle = - (0.24 \pm 0.01 \,\text{GeV})^3, \quad \langle \bar{q}g_s \sigma G q \rangle = m_0^2 \langle \bar{q}q \rangle, \quad \langle \bar{s}s \rangle = (0.8 \pm 0.1) \langle \bar{q}q \rangle, \quad \langle \bar{s}g_s \sigma G s \rangle = m_0^2 \langle \bar{s}s \rangle, \quad m_0^2 = (0.8 \pm 0.1) \,\text{GeV}^2, \quad \langle \frac{\alpha_s GG}{\pi}\rangle=(0.33 \,\text{GeV})^4$\cite{ref102,ref103,ref104}.
For the masses of the $c/b/s$ quarks, we use the $\overline{MS}$ masses: $m_c(m_c) = (1.275 \pm 0.025) \,\text{GeV}, m_b(m_b) = (4.18 \pm 0.03) \,\text{GeV}, m_s(\mu=2\,\text{GeV}) = (0.095 \pm 0.005) \,\text{GeV} $. These masses evolve with the energy scale $\mu$ according to the renormalization group equations\cite{ref122}:

\begin{equation*}
\langle \bar{q}q \rangle (\mu) = \langle \bar{q}q \rangle (1\text{GeV}) \left[ \frac{\alpha_s(1\text{GeV})}{\alpha_s(\mu)} \right]^{\frac{12}{33 - 2n_f}} ,
\end{equation*}
\begin{equation*}
\langle \bar{s}s \rangle (\mu) = \langle \bar{s}s \rangle (1\text{GeV}) \left[ \frac{\alpha_s(1\text{GeV})}{\alpha_s(\mu)} \right]^{\frac{12}{33 - 2n_f}} ,
\end{equation*}
\begin{equation*}
\begin{aligned}
    \langle \bar{q}g_s \sigma G q \rangle (\mu) & = \langle \bar{q}g_s \sigma G q \rangle (1\text{GeV}) \\
    & \times \left[ \frac{\alpha_s(1\text{GeV})}{\alpha_s(\mu)} \right]^{\frac{2}{33 - 2n_f}} ,
\end{aligned}
\end{equation*}
\begin{equation*}
\begin{aligned}
    \langle \bar{s}g_s \sigma G s \rangle (\mu) &= \langle \bar{s}g_s \sigma G s \rangle (1\text{GeV}) \\
    & \times \left[ \frac{\alpha_s(1\text{GeV})}{\alpha_s(\mu)} \right]^{\frac{2}{33 - 2n_f}} ,
\end{aligned}
\end{equation*}
\begin{equation*}
m_Q(\mu) = m_Q(m_Q) \left[ \frac{\alpha_s(\mu)}{\alpha_s(m_Q)} \right]^{\frac{12}{33 - 2n_f}} ,
\end{equation*}
\begin{equation*}
m_s(\mu) = m_s(2\text{Gev}) \left[ \frac{\alpha_s(\mu)}{\alpha_s(2 \text{Gev})} \right]^{\frac{12}{33 - 2n_f}} ,
\end{equation*}
\begin{equation}
\begin{aligned}
\alpha_s(\mu) & =  \frac{1}{b_0 t}  \left[  1 - \frac{b_1}{b_0^2} \frac{\log t}{t} \right. \\
& \left.  + \frac{b_1^2 (\log^2 t - \log t - 1) + b_0 b_2}{b_0^4 t^2} \right] \,.
\label{Eq.(14)}
\end{aligned}
\end{equation}
where \( t = \log\frac{\mu^2}{\Lambda_{\text{QCD}}^2} \),  \( b_0 = \frac{33 - 2n_f}{12\pi} \), \( b_1 = \frac{153 - 19n_f}{24\pi^2} \), \( b_2 = \frac{2857 - \frac{5033}{9} n_f + \frac{325}{27} n_f^2}{128\pi^3} \). The number of active quark flavors \( n_f \) takes the values \( 5,\,4,\, \) and \(\,3 \), with the corresponding \( \Lambda_{\text{QCD}} \) values being \( 213 \, \text{MeV}, 296 \, \text{MeV}, \) and \( 339 \, \text{MeV} \), respectively\cite{ref93}. For the studies of \(cc\bar{c}\bar{q}\) and \(cc\bar{c}\bar{s}\), we take \(n_f=4\), and for the studies of \(bb\bar{b}\bar{q}\) and \(bb\bar{b}\bar{s}\), we take \(n_f=5\).

\begin{figure*}
\centering
\begin{minipage}{0.5\textwidth}
  \centering
  \includegraphics[width=85mm]{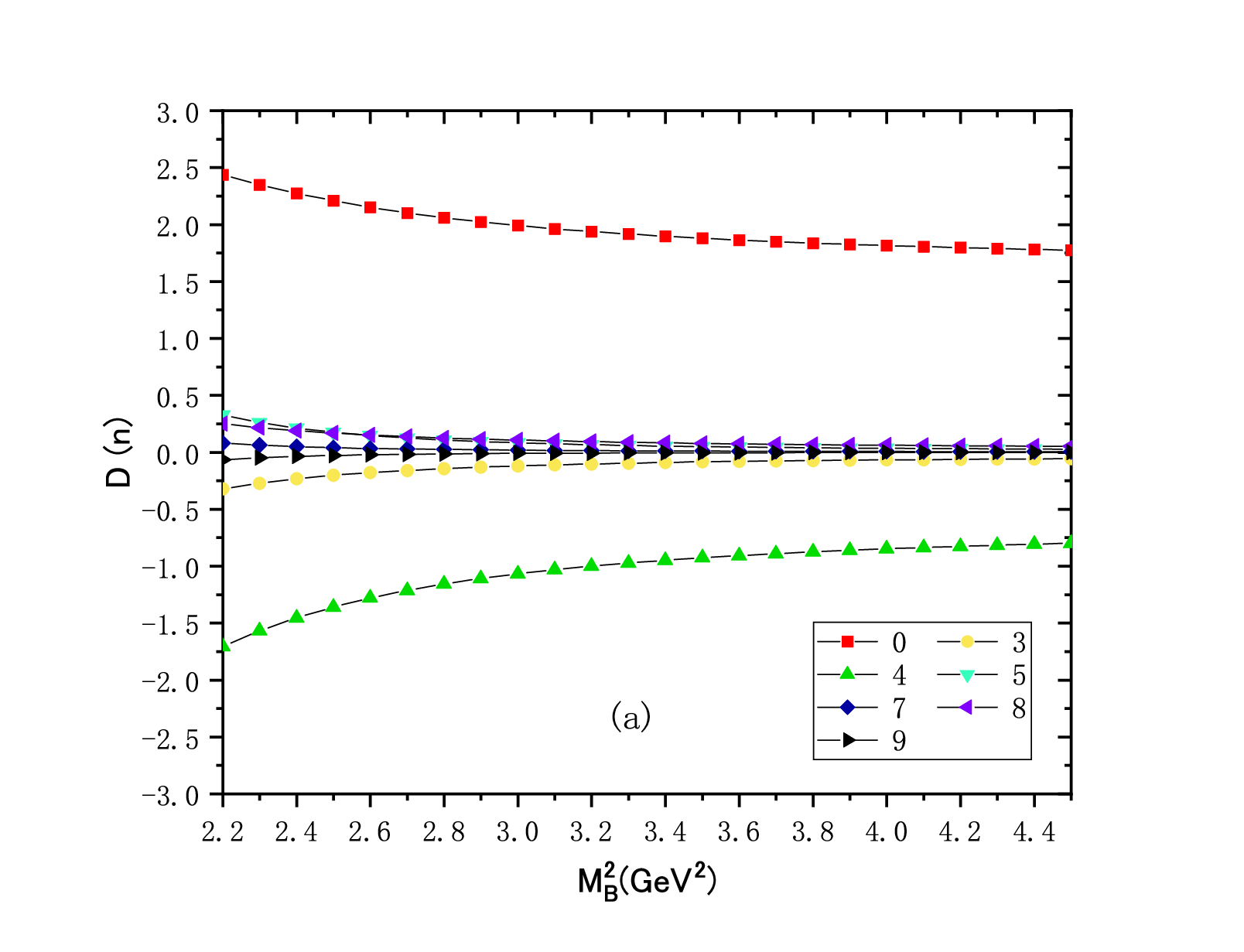}
  \label{fig:one_a}
\end{minipage}%
\begin{minipage}{0.5\textwidth}
  \centering
  \includegraphics[width=85mm]{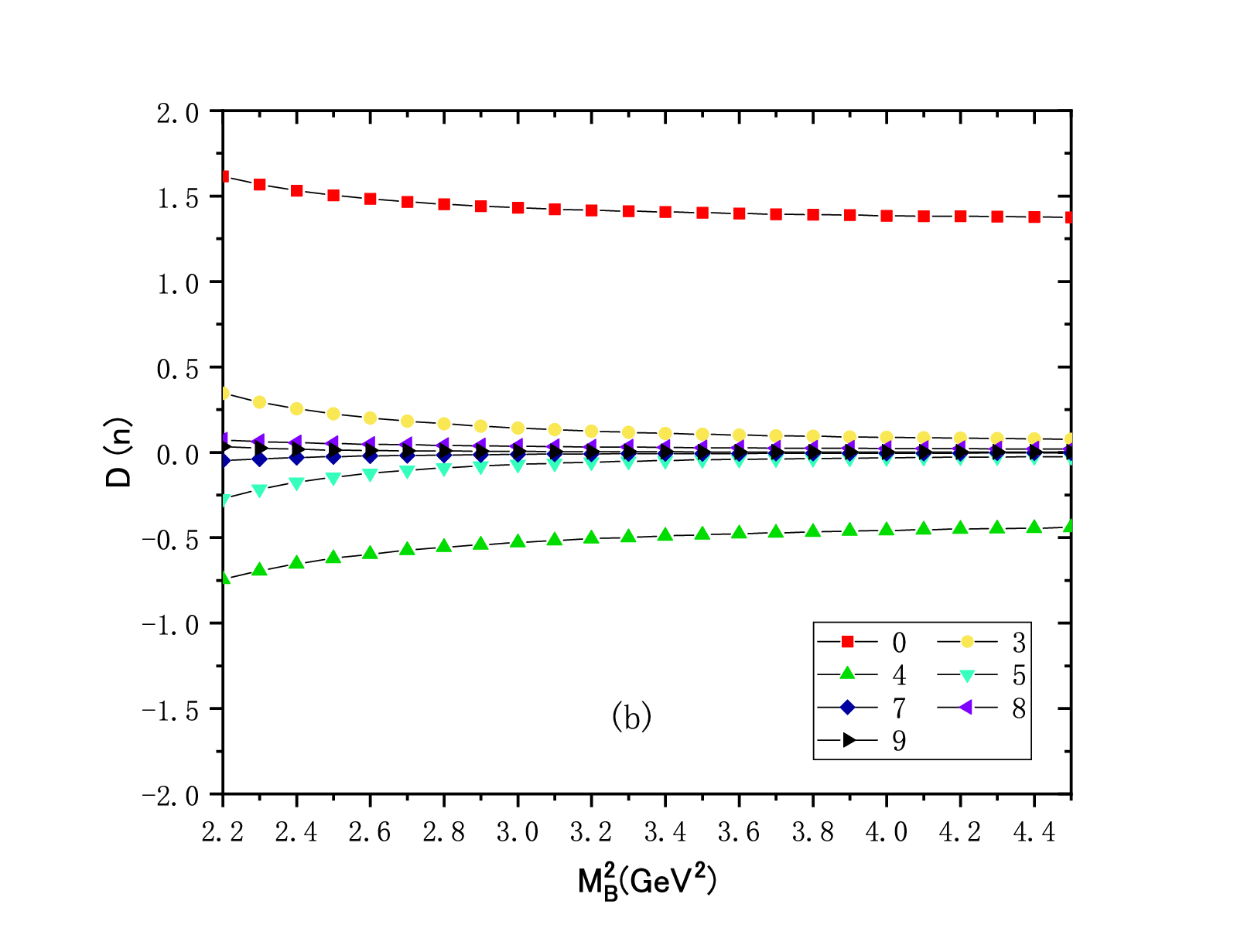}
  \label{fig:one_b}
\end{minipage} \\[0.1cm] 
\begin{minipage}{0.5\textwidth}
  \centering
  \includegraphics[width=85mm]{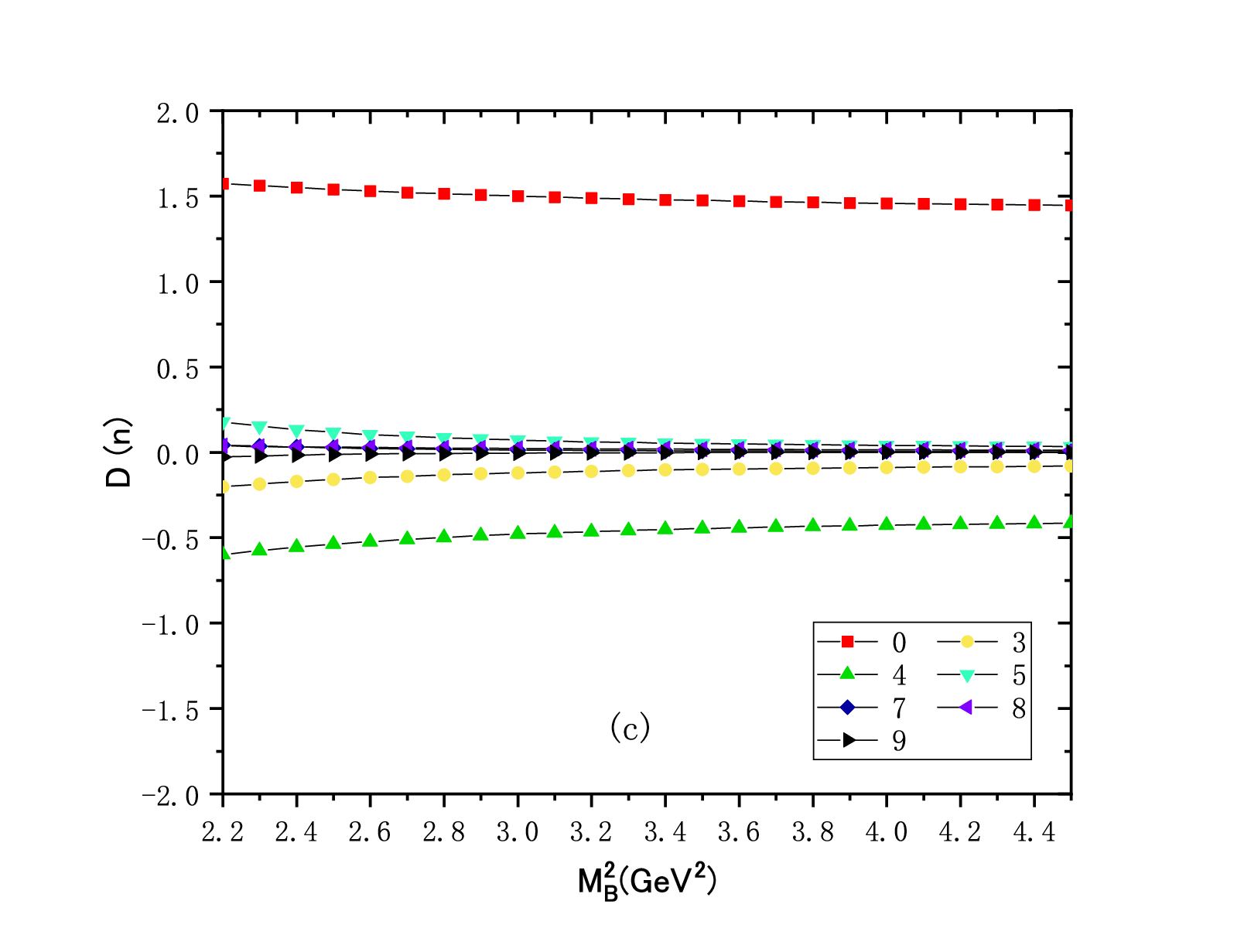}
  \label{fig:one_c}
\end{minipage}%
\begin{minipage}{0.5\textwidth}
  \centering
  \includegraphics[width=85mm]{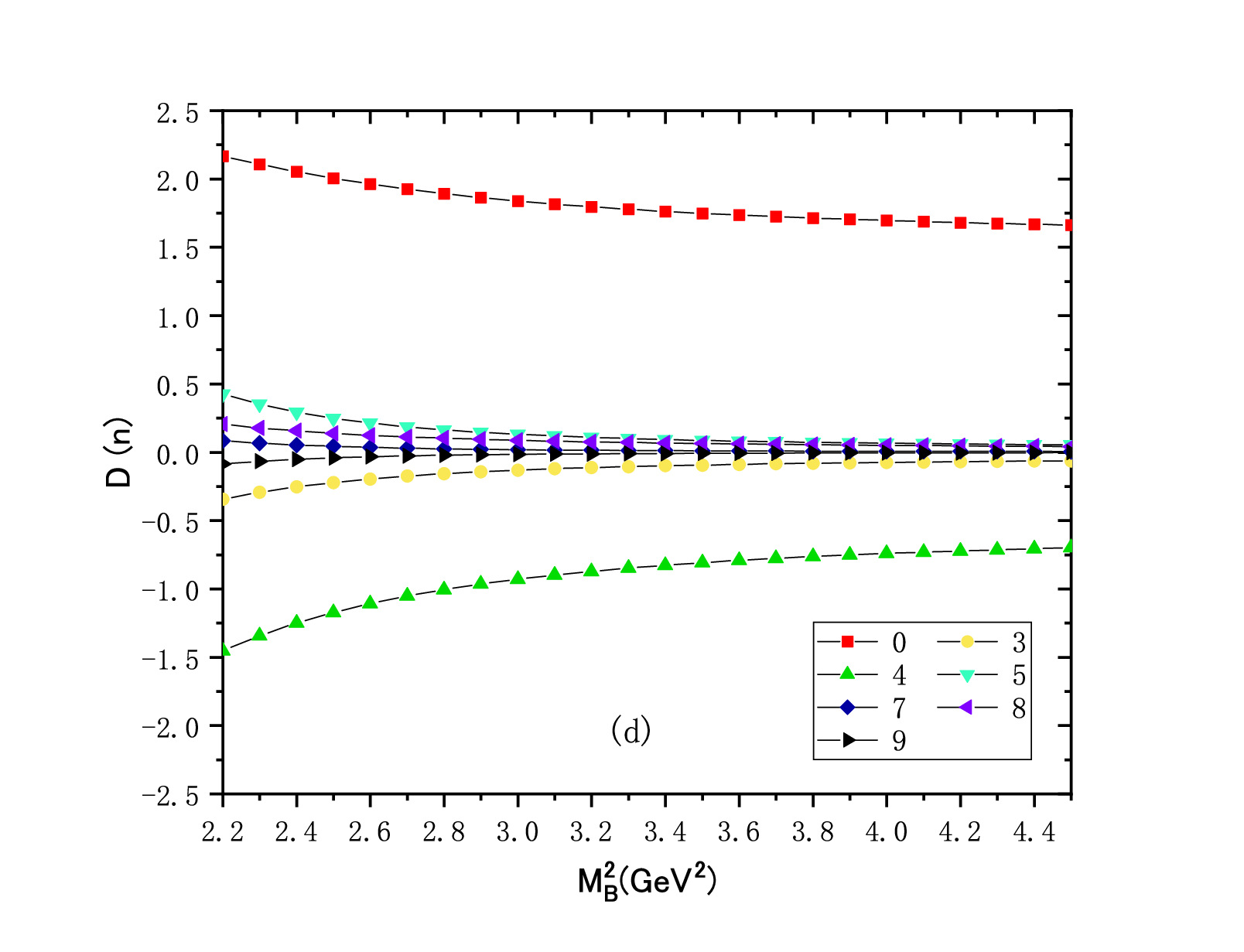}
  \label{fig:one_d}
\end{minipage}
  \label{Fig.1}
\caption{The contributions of \(D(n)\) in the operator product expansion varies with the parameter \(M_B^2\). Figures (a), (b), (c), and (d) correspond to \( J_1^{cc\bar{c}\bar{q}} \), \( J_4^{cc\bar{c}\bar{q}} \), \( J_5^{cc\bar{c}\bar{q}} \), and \( J_7^{cc\bar{c}\bar{q}} \), with threshold parameters \( \sqrt{s_0} \) of 6.10 GeV, 6.10 GeV, 5.90 GeV, and 6.20 GeV, respectively.}
\end{figure*}

\begin{table*}
\centering
\caption{Parameters for the scalar tetraquark states: The table summarizes the values of \(M_B^2\), threshold parameters, energy scales, pole contributions, masses, and the contributions from the dimension$\text{-9}$ term in the operator product expansion.
 }
\begin{tabular}{lcccccccc}
\toprule
\textbf{}&\textbf{$QQ\bar{Q}\bar{q}$} & \textbf{$J^{P}$} & \textbf{$M_B^2 (\text{GeV}^{2})$} & \textbf{$\sqrt{s_0} (\text{GeV})$} & \textbf{$\mu (\text{GeV})$} & \textbf{Pole (\%)}&\textbf{Mass(GeV)}&  \textbf{$|D(9)|$} \\
\midrule
$J_1$ &$[cc]_S[\bar{c}\bar{q}]_S$ & $0^{+}$ & $3.1-3.5$ & $6.10 \pm 0.10$ & 1.8 & $40 - 58$ &$5.66 \pm 0.09$& $0.95\%$ \\
$J_4$&$[cc]_A[\bar{c}\bar{q}]_A$ & $0^{+}$ & $2.7-3.2$ & $6.10 \pm 0.10$ & 1.3 & $40 - 57$ &$5.66 \pm 0.10 $& $0.98\%$ \\
$J_5$&$[cc]_V[\bar{c}\bar{q}]_V$ & $0^{+}$ & $2.5-2.9$ & $5.90 \pm 0.10$ & 1.3
& $39 - 59$ &$5.50\pm 0.12$& $0.97\%$ \\
$J_7$&$[cc]_P[\bar{c}\bar{q}]_P$ & $0^{+}$ & $3.2-3.7$ & $6.20 \pm 0.10$ & 1.8 & $39 - 59$ &$5.73 \pm 0.11$& $0.97\%$ \\

\midrule
$J_1$ &$[cc]_S[\bar{c}\bar{s}]_S$ & $0^{+}$ & $3.7-4.3$ & $6.50 \pm 0.10$ &2.0 & $39 - 58$ &$5.95 \pm 0.12$& $0.95\%$ \\
$J_4$&$[cc]_A[\bar{c}\bar{s}]_A$ & $0^{+}$ & $3.4-3.9$ & $6.30 \pm 0.10$ & 1.8 & $40 - 59$ &$5.78 \pm 0.12$& $0.97\%$ \\
$J_5$&$[cc]_V[\bar{c}\bar{s}]_V$ & $0^{+}$ & $3.4-3.9$ & $6.30 \pm 0.10$ & 1.8
& $40 - 60$ &$5.79\pm 0.12$& $0.98\%$ \\
$J_7$&$[cc]_P[\bar{c}\bar{s}]_P$ & $0^{+}$ & $3.9-4.5$ & $6.60 \pm 0.10$ &2.1 & $39 - 59$ &$6.04 \pm 0.12$& $0.98\%$ \\

\midrule
$J_1$ &$[bb]_S[\bar{b}\bar{q}]_S$ & $0^{+}$ & $8.6-9.5$ & $15.80 \pm 0.10$ & 2.5& $41 - 59$ &$15.29\pm 0.17$& $ 0.99\%$ \\
$J_4$&$[bb]_A[\bar{b}\bar{q}]_A$ & $0^{+}$ & $8.5-10.8$ & $15.90 \pm 0.10$ & 2.5 & $40 - 59$ &$15.29 \pm 0.17$ & $\ll 1\%$ \\
$J_5$&$[bb]_V[\bar{b}\bar{q}]_V$ & $0^{+}$ & $7.0-8.4$ & $15.50 \pm 0.10$ & 2.5
& $40 - 60$ &$15.02\pm 0.13$& $\ll 1\%$ \\
$J_7$&$[bb]_P[\bar{b}\bar{q}]_P$ & $0^{+}$ & $8.6-9.5$ & $15.90 \pm 0.10$ & 2.5& $39 - 59$ &$15.38\pm 0.17$& $0.97\%$ \\

\midrule
$J_1$ &$[bb]_S[\bar{b}\bar{s}]_S$ & $0^{+}$ & $8.7-9.5$ & $15.85 \pm 0.10$ & 2.5 & $41 - 59$ &$15.38 \pm 0.17$& $0.99\%$ \\
$J_4$&$[bb]_A[\bar{b}\bar{s}]_A$ & $0^{+}$ & $8.7-11.0$ & $15.95 \pm 0.10$ & 2.5  & $40 - 59$ &$15.38 \pm 0.17$& $\ll 1\%$ \\
$J_5$&$[bb]_V[\bar{b}\bar{s}]_V$ & $0^{+}$ & $7.3-8.6$ & $15.55\pm 0.10$ & 2.5
& $40 - 60$ &$15.08\pm 0.13$& $0.96\%$ \\
$J_7$&$[bb]_P[\bar{b}\bar{s}]_P$ & $0^{+}$ & $9.0-10.0$ & $15.95 \pm 0.10$ & 2.5 & $39 - 58$ &$ 15.49 \pm 0.17 $& $
0.99\%$ \\

\bottomrule
\end{tabular}
\label{TableI}
\end{table*}

Refs. \cite{ref89, ref90, ref91, ref92} proposed a crucial energy scale formula that played a significant role in the mass spectrum studies of hidden charm tetraquark states \cite{ref101, ref105, ref106} and triply charm pentaquark states \cite{ref107}. Both hidden charm tetraquark states and triply heavy tetraquark states belong to the tetraquark states, while triply heavy tetraquark states and triply heavy pentaquark states include three heavy quarks in their quark compositions. Hence, there are grounds to believe that this energy scale formula may be applicable to the study of triply heavy tetraquark states. Therefore, we used the energy scale formula to test our predictions for the masses of triply heavy tetraquark states. We have abbreviated the energy scale formula for ease of application:

\begin{equation}
    \mu = \sqrt{M_{X}^2 - (i\mathbb{M}_c + j\mathbb{M}_b)^2}-k\mathbb{M}_s \,\,,
\label{Eq.(15)}
\end{equation}
where, \(i\) represents the number of \(c\) quarks in the interpolating currents in Eq. (\hyperref[Eq.(1)]{1}), \(j\) represents the number of \(b\) quarks, and \(k\) represents the number of \(s\) quarks. In the current work, we employ updated effective masses for the \(c\), \(b\), and \(s\) quarks \cite{ref108,ref109}:
\begin{equation}
\mathbb{M}_c = 1.82  \,\text{GeV},\ \mathbb{M}_b = 5.17 \,\text{GeV},\ \mathbb{M}_s = 0.20 \, \text{GeV} \,.
\label{Eq.(16)}
\end{equation}

When choosing \( M_B^2 \) and \( s_0 \), two criteria must be met. The first ensures the phenomenological part is dominated by the pole contribution (PC), defined as:
\begin{equation}
\text{PC} = \frac{\int_{9m_Q^2}^{s_0} ds \, \rho_{\text{OPE}}(s) e^{-\frac{s}{M_B^2}}}{\int_{9m_Q^2}^{\infty} ds \,  \rho_{\text{OPE}}(s) e^{-\frac{s}{M_B^2}}} \,.
\label{Eq.(17)}
\end{equation}

The pole dominance criterion requires that the pole contribution (PC) to be within the range of \(40\%-60\%\). The second criterion concerns the convergence of the operator product expansion (OPE). To assess the convergence of the OPE, we calculate the contributions from the vacuum condensate terms of various dimensions within the OPE, denoted as \(D(n)\), using the following formula:
\begin{equation}
D(n) = \frac{\int_{9m_Q^2}^{s_0} ds \, \rho^{(n)}_{\text{OPE}}(s) e^{-\frac{s}{M_B^2}}}{\int_{9m_Q^2}^{s_0} ds \, \rho_{\text{OPE}}(s) e^{-\frac{s}{M_B^2}}}\, .
\label{Eq.(18)}
\end{equation}

Here, \(\rho^{(n)}_{\text{OPE}}(s)\) denotes the term in \(\rho_{\text{OPE}}(s)\) corresponding to the vacuum condensate of dimension \(n\). We impose the condition that the contribution of the vacuum condensate with dimension 9 (the highest dimension) satisfies \(|D(9)| \leq 1\%\), and use this as a constraint to select the parameter \( M_B^2 \) and the threshold parameter \( s_0 \).

\begin{figure*}
\centering
\begin{minipage}{0.5\textwidth}
  \centering
  \includegraphics[width=65mm]{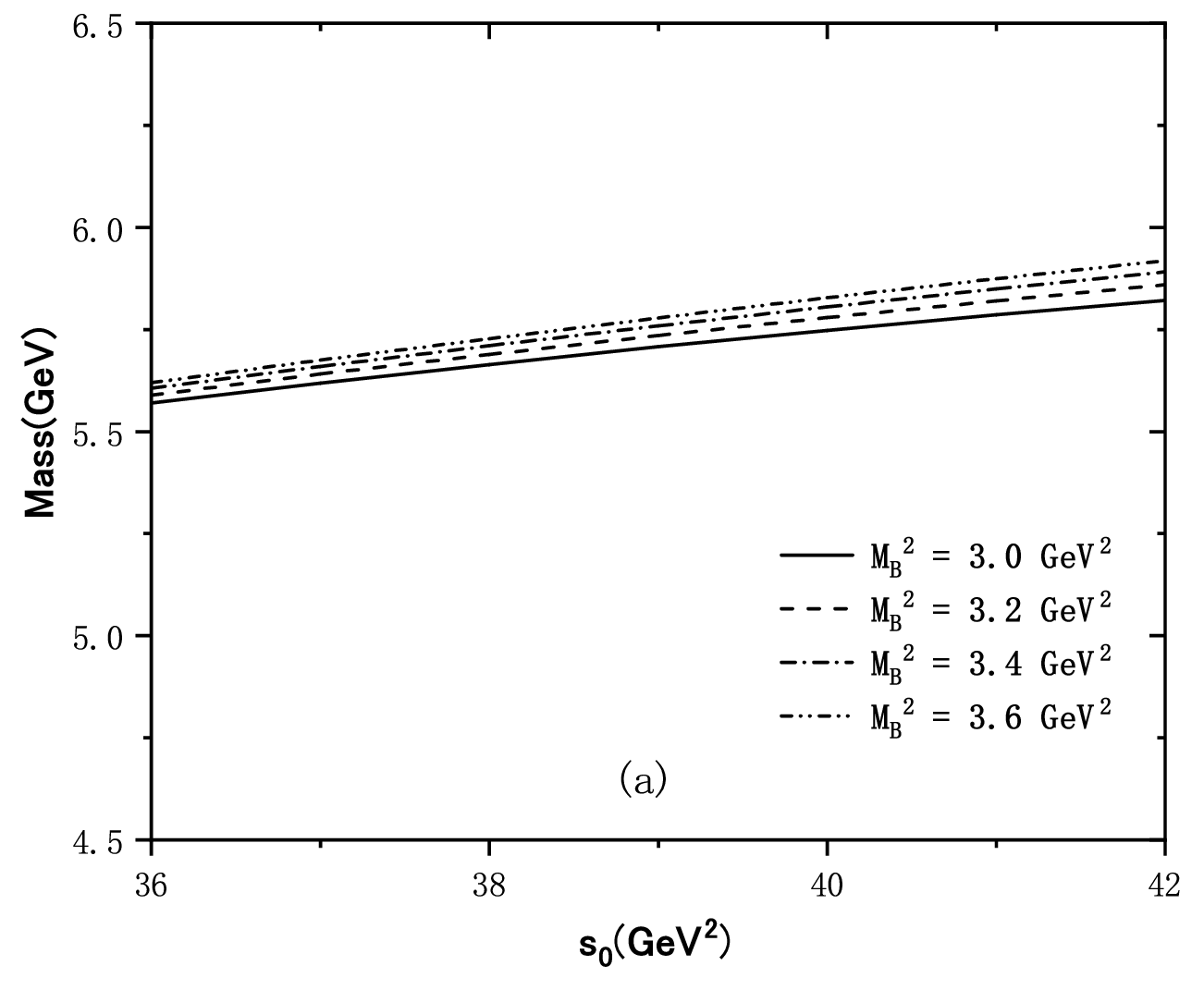}
  \label{fig:two_a}
\end{minipage}%
\begin{minipage}{0.5\textwidth}
  \centering
  \includegraphics[width=65mm]{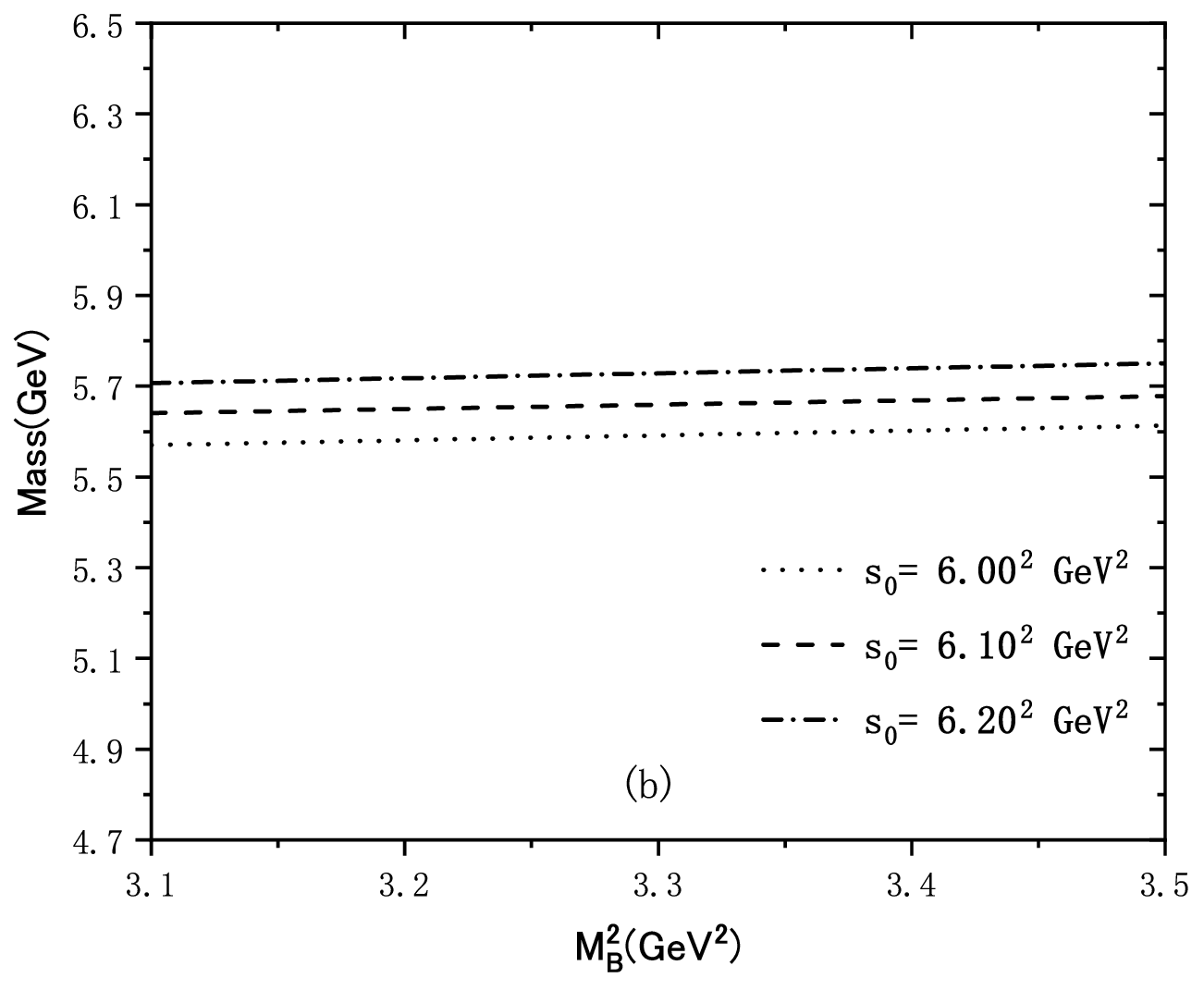}
  \label{fig:two_b}
\end{minipage} \\[0.1cm] 
\begin{minipage}{0.5\textwidth}
  \centering
  \includegraphics[width=65mm]{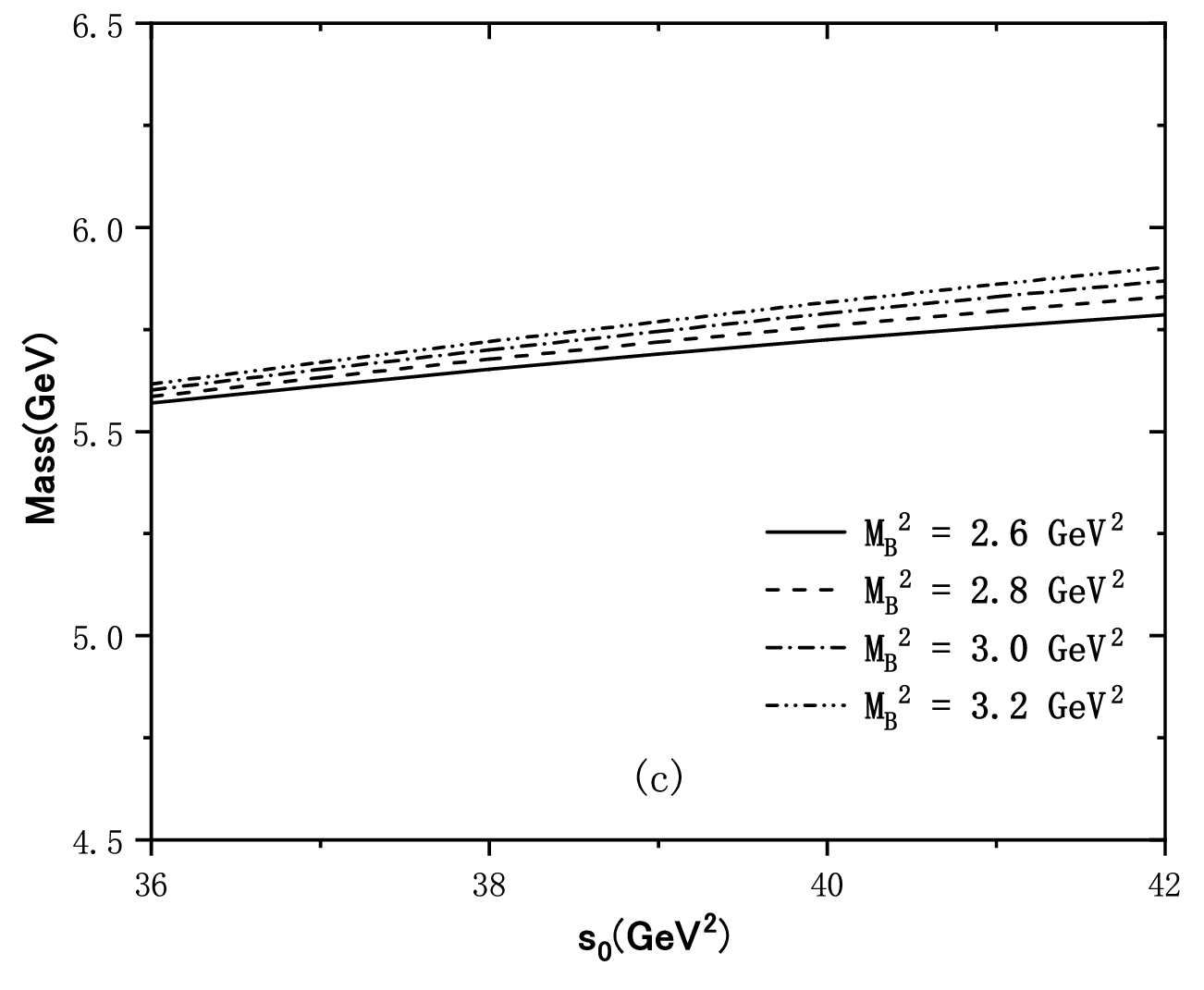}
  \label{fig:two_c}
\end{minipage}%
\begin{minipage}{0.5\textwidth}
  \centering
  \includegraphics[width=65mm]{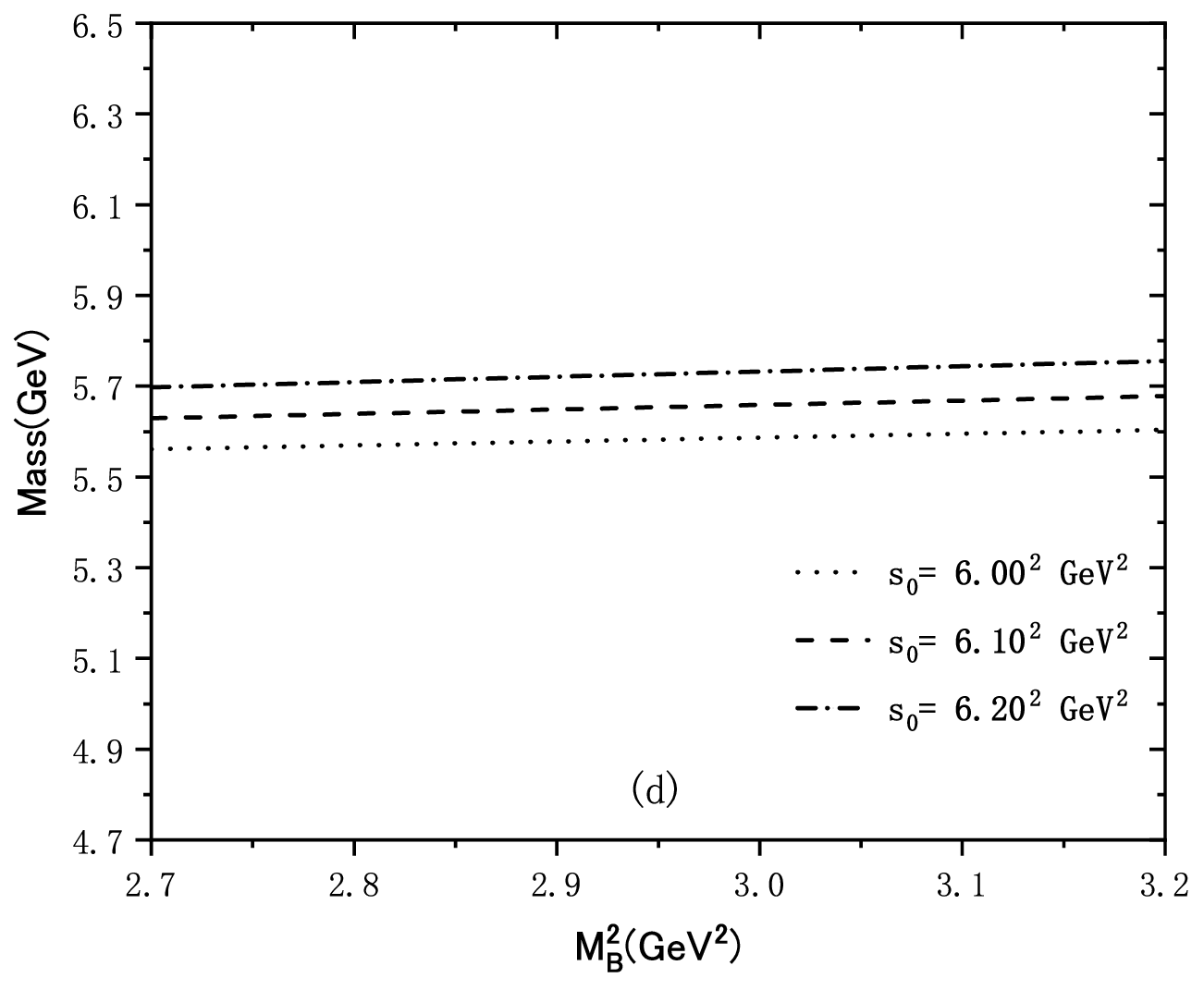}
  \label{fig:two_d}
\end{minipage}
 \label{Fig.2}
\caption{Mass dependence on parameters \(s_0\) and \(M_B^2\) for $J_1^{cc\bar{c}\bar{q}}$ and $J_4^{cc\bar{c}\bar{q}}$: (a) and (b) depict the mass variation with \(s_0\) and \(M_B^2\) for $J_1^{cc\bar{c}\bar{q}}$, respectively; (c) and (d) illustrate the corresponding variations for $J_4^{cc\bar{c}\bar{q}}$.} 
\end{figure*}

\begin{figure*}
\centering
\begin{minipage}{0.5\textwidth}
  \centering
  \includegraphics[width=65mm]{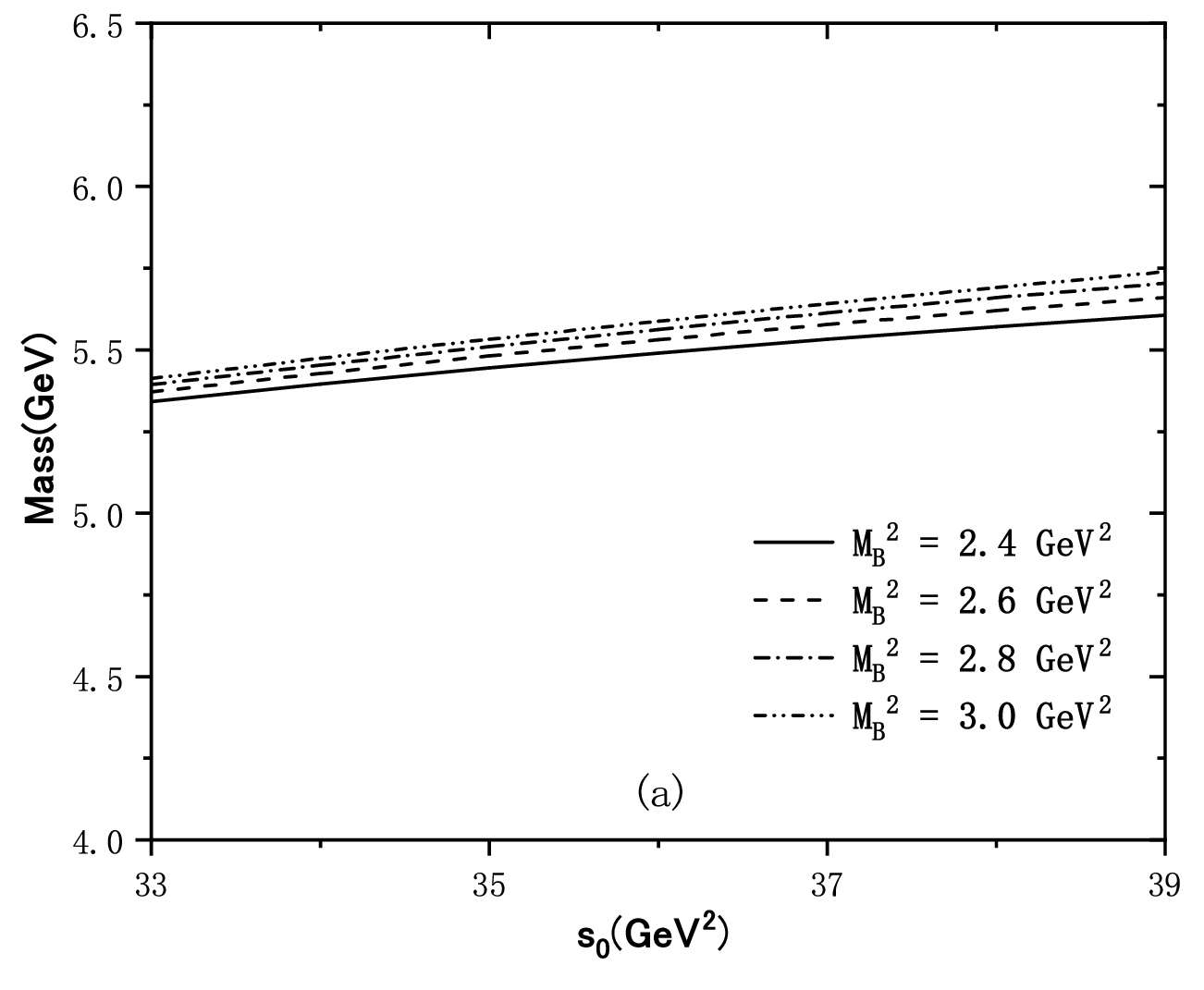}
  \label{fig:three_a}
\end{minipage}%
\begin{minipage}{0.5\textwidth}
  \centering
  \includegraphics[width=65mm]{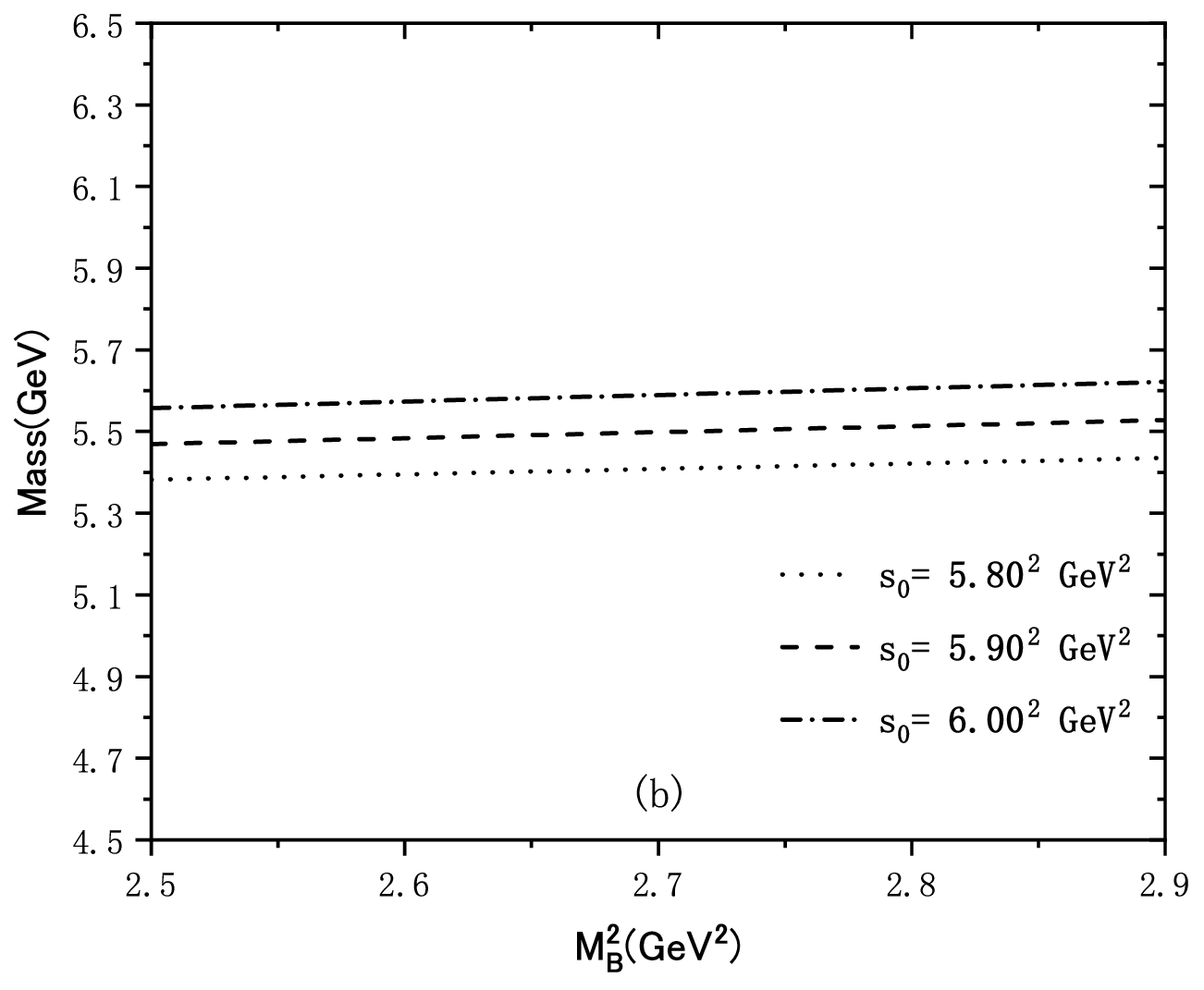}
  \label{fig:three_b}
\end{minipage} \\[0.1cm] 
\begin{minipage}{0.5\textwidth}
  \centering
  \includegraphics[width=65mm]{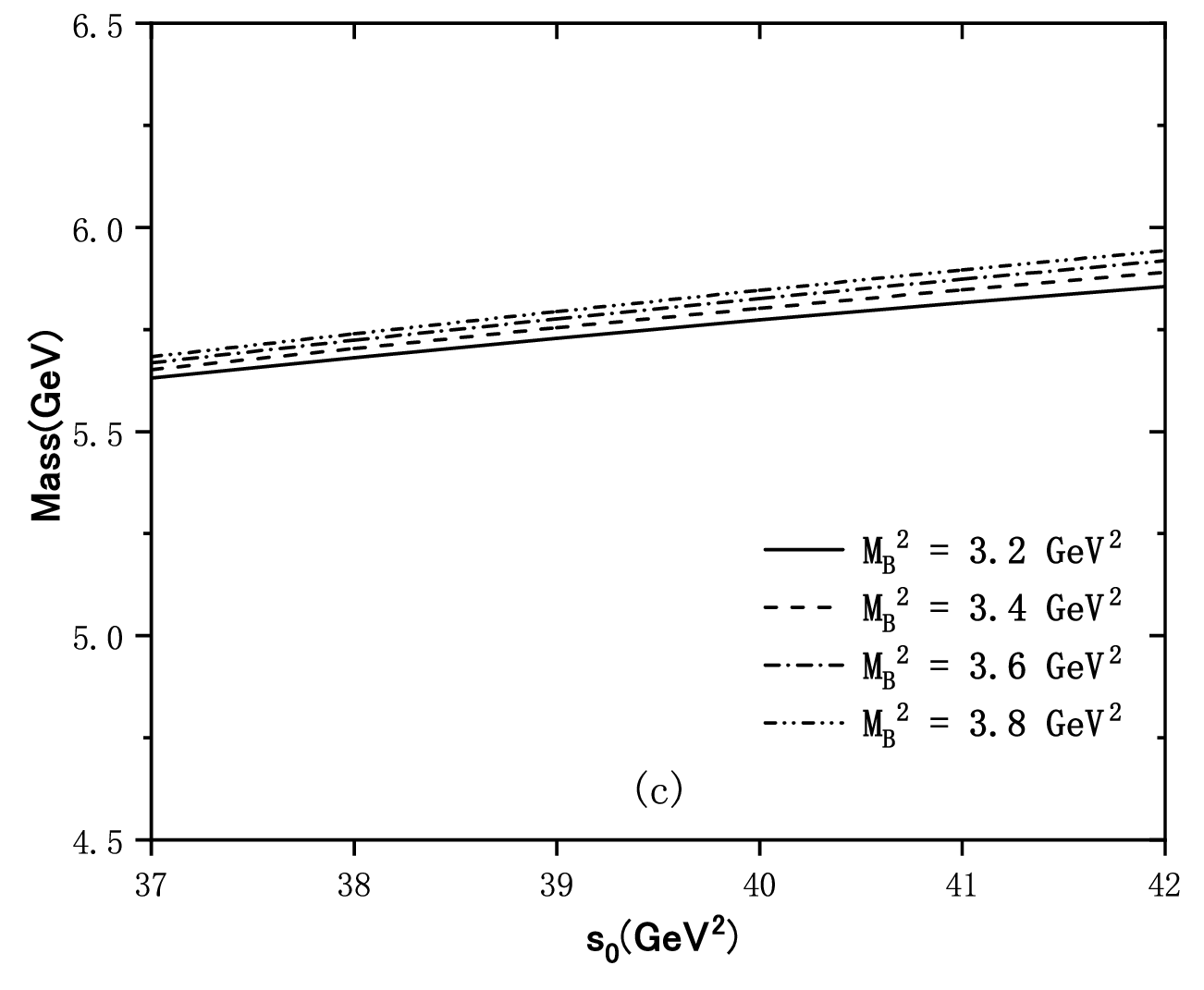}
  \label{fig:three_c}
\end{minipage}%
\begin{minipage}{0.5\textwidth}
  \centering
  \includegraphics[width=65mm]{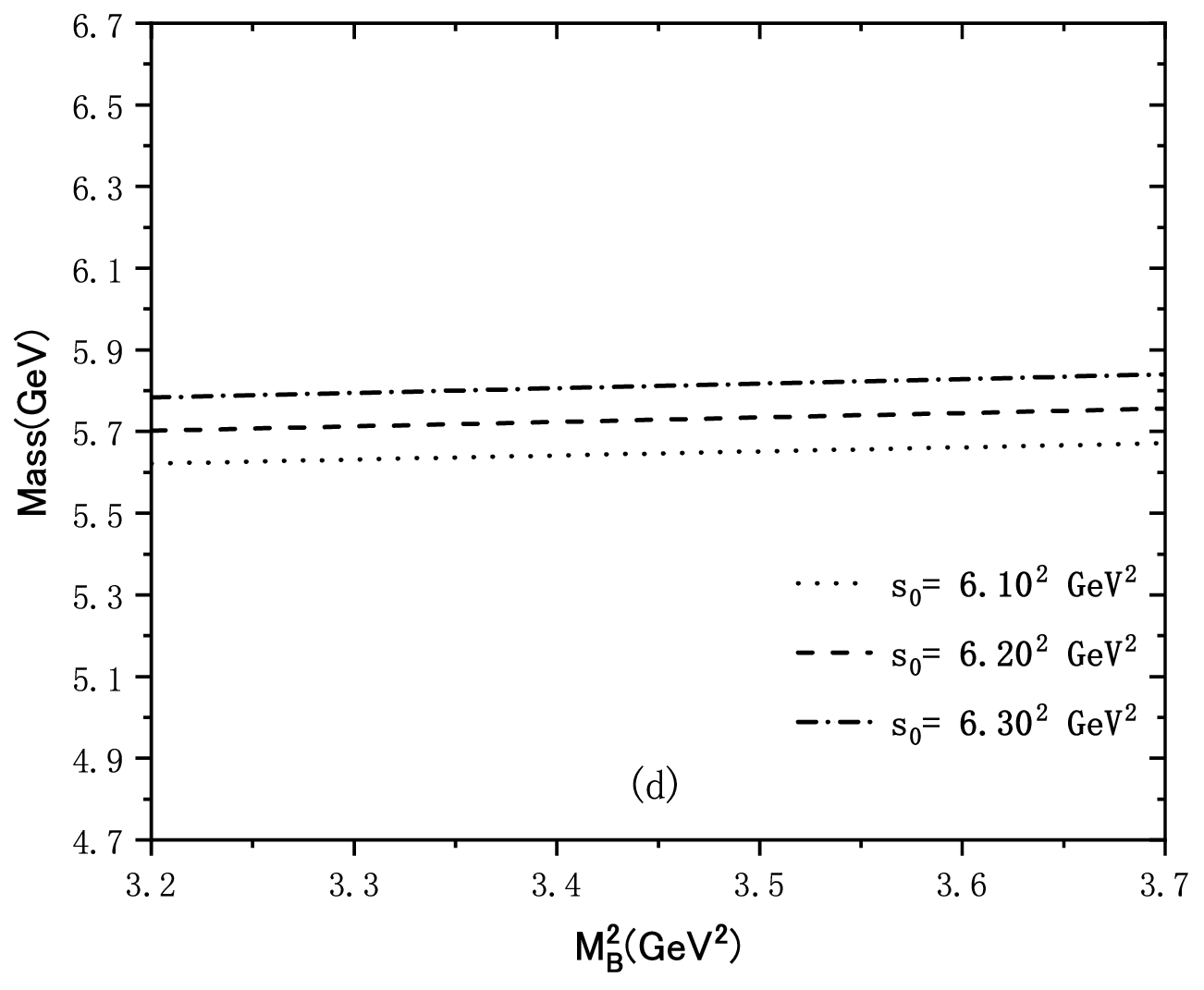}
  \label{fig:three_d}
\end{minipage}
 \label{Fig.3}
\caption{The same caption as in Fig.2, but for $J_5^{cc\bar{c}\bar{q}}$ and $J_7^{cc\bar{c}\bar{q}}$, respectively. }
\end{figure*}

Next, we will provide a detailed description of our numerical analysis process. The first step is to choose a scale parameter $\mu$ within a reasonable range. In our study, we found that for the numerical analysis of $cc\bar{c}\bar{q}$ and $cc\bar{c}\bar{s}$ systems, the range of $\mu$ should not exceed $1.3-2.1\,\text{GeV}$ , while, for the analysis of $bb\bar{b}\bar{q}$ and $bb\bar{b}\bar{s}$ systems, the range of $\mu$ should not exceed $2.5-2.9\,\text{GeV}$. If the value of $\mu$ exceeds these ranges, it is not possible to obtain results that satisfy both criteria—the dominance of the pole contribution (PC) and the convergence of the operator product expansion (OPE)—while also ensuring a sufficiently smooth Borel platform. Similar discussions can be found in Ref.\cite{ref121}. The second step is to introduce the choice of the Borel parameter $M_B^2$. In general, when a tentative  threshold parameter \( s_0 \) is given, the \text{PC} decreases monotonically with the increase of \( M_B^2 \). We define the minimum \( M_B^2 \) as the value corresponding to $\text{PC} = 60\%$, and the maximum \( M_B^2 \) as the value corresponding to $\text{PC} = 40\%$. The third step in our study is that we found the threshold \(s_0\) cannot be too small. When \(s_0\) is too small, the pole contribution (PC) cannot reach 60\%, and the value of \(|D(9)|\) within the Borel window fails to meet the requirement of being less than or equal to 1\%. Therefore, in order to satisfy both criteria, excessively small values of $s_0$ should be discarded. After selecting the threshold parameter $s_0$, we obtain the minimum value of $s_0$. At the selected threshold parameter \(s_0\), we analyze how the contributions of \(D(n)\) in the operator product expansion vary with the parameter \(M_B^2\). As shown in Fig. \hyperref[Fig.1]{1}, figures (a), (b), (c), and (d) correspond to \( J_1^{cc\bar{c}\bar{q}} \), \( J_4^{cc\bar{c}\bar{q}} \), \( J_5^{cc\bar{c}\bar{q}} \), and \( J_7^{cc\bar{c}\bar{q}} \), respectively, with threshold parameters \( \sqrt{s_0} \) of 6.10 GeV, 6.10 GeV, 5.90 GeV, and 6.20 GeV. Meanwhile, we plot functions showing how the hadron mass changes as a function of $s_0$ and $M_B^2$, as shown in Figs. \hyperref[Fig.2]{2} and \hyperref[Fig.3]{3}. A reasonable physical result requires very little variation with changes in the model parameters. That is, the hadron mass should vary very little with $M_B^2$ near a certain $s_0$ point. By analyzing functions showing how the hadron mass depends on $s_0$ and the Borel parameter $M_B^2$, we can determine the optimal threshold parameter $s_0$. The fourth step is to substitute the chosen scale parameter $\mu$, the optimal threshold parameter $s_0$, and the corresponding Borel parameter $M_B^2$ into Eq. (\hyperref[Eq.(13)]{13}) to extract the hadron mass. If the extracted hadron mass does not satisfy Eq. (\hyperref[Eq.(15)]{15}) for the chosen scale $\mu$, we need to adjust the value of $\mu$ and repeat the above steps until the extracted hadron mass and the scale $\mu$ satisfy Eq. (\hyperref[Eq.(15)]{15}).

After completing the above numerical analysis, we can obtain the hadron masses for the currents in Eq. (\hyperref[Eq.(1)]{1}) via Eq. (\hyperref[Eq.(13)]{13}), as summarized in Table \hyperref[TableI]{I}. This table also includes the working interval for \(M_B^2\), the values of \(s_0\) and \(\mu\), the pole contributions (PC), and the contributions from the dimension$\text{-9}$ term in the operator product expansion (\(|D(9)|\)). Regarding the uncertainties in hadron mass, we take into account not only the errors from \(s_0\) and \(M_B^2\), but also those from the input parameters.

In the analysis of the triply bottom tetraquark states, we found that within the reasonable range of \(\mu\), adjusting its value does not lead to results that satisfy Eq. (\hyperref[Eq.(15)]{15}). Based on our numerical analysis presented in Table \hyperref[TableI]{I}, we suggest that for the analysis of triply bottom tetraquark states, a value of \(\mathbb{M}_b = 5.05 \, \text{GeV}\) for the effective mass of the bottom quark should be adopted, as indicated in the analysis.

\section{CONCLUSIONS}
\label{sec:CONCLUSIONS}

In this article, we analyze the types \(C\gamma_5 \otimes \gamma_5 C\), \(C\gamma_\mu \otimes \gamma^\mu C\), \(C\gamma_\mu \gamma_5 \otimes \gamma_5 \gamma^\mu C\), and \(C \otimes C\) of triply heavy tetraquark states with quantum number $J^P=0^+$, namely \(cc\bar{c}\bar{q}\), \(cc\bar{c}\bar{s}\), \(bb\bar{b}\bar{q}\), and \(bb\bar{b}\bar{s}\), using the QCD sum rules. This is achieved by calculating the contributions from vacuum condensates up to dimension-9 in the operator product expansion. During numerical calculations, we use the energy scale formula $\mu = \sqrt{M_{X}^2 - (i\mathbb{M}_c + j\mathbb{M}_b)^2} - k\mathbb{M}_s$ to determine the optimal energy scale for the QCD spectral density, following the two criteria of the QCD sum rules to select the ideal Borel parameter $M_B^2$ and the threshold parameter $s_0$.

\begin{table}[htbp]
\renewcommand{\arraystretch}{1.5} 
\setlength{\tabcolsep}{2.9pt}
\centering
\caption{Comparison of our predicted masses (GeV) for triply heavy tetraquark states with quantum number \(0^+\) with those from other theoretical works.
}
\begin{tabular}{lcccccccc}
\toprule
\textbf{}&\textbf{$cc\bar{c}\bar{q}$} & $cc\bar{c}\bar{s}$ & $bb\bar{b}\bar{q}$  & $bb\bar{b}\bar{s}$  \\
\midrule
$\text{This work}$& $5.61_{-0.23}^{+0.23}
$ & $5.91_{-0.25}^{+0.25}
$ & $15.22_{-0.33}^{+0.33}
$ & $15.30_{-0.35}^{+0.36}
$  \\
\,\,\,\,\,\,\,\,\cite{ref78}& $5.71_{-0.12}^{+0.12}
$ & $5.77_{-0.09}^{+0.09}
$ & $15.36_{-0.04}^{+0.05}
$ & $15.45_{-0.07}^{+0.08}
$  \\
\,\,\,\,\,\,\,\,\cite{ref85}& $5.10_{-0.20}^{+0.20}
$ & $-
$ & $13.50_{-0.4}^{+0.4}
$ & $-
$  \\
\,\,\,\,\,\,\,\,\cite{ref84} & $5.40
$ & $5.48
$ & $15.11
$ & $15.18
$  \\

\bottomrule
\end{tabular}
\label{TableII}
\end{table}

As a result, we obtain the masses of various scalar tetraquark states. Our findings show that the triply charm tetraquark states \(cc\bar{c}\bar{q}\) and \(cc\bar{c}\bar{s}\) lie within the ranges of $5.38-5.84\,\text{GeV}$ and $5.66-6.16\,\text{GeV}$, respectively. In the bottom sector, the triply bottom tetraquark states \(bb\bar{b}\bar{q}\) and \(bb\bar{b}\bar{s}\) have ranges of $14.89-15.55\,\text{GeV}$ and $14.95-15.66\,\text{GeV}$, respectively. As shown in Table \hyperref[TableII]{II}, our results agree with the theoretical predictions of Refs. \cite{ref78,ref84}, within the respective error margins.

Moreover, each of the tetraquark states \( cc\bar{c}\bar{q} \), \( cc\bar{c}\bar{s} \), \( bb\bar{b}\bar{q} \), and \( bb\bar{b}\bar{s} \) can be classified into four types according to the diquark fields. For the \( cc\bar{c}\bar{q} \) state, as an example, the mass ranges for different configurations are as follows: for the \( C\gamma_5 \otimes \gamma_5 C \) type, the mass is between \( 5.57 - 5.75 \, \text{GeV} \); for the \( C\gamma_\mu \otimes \gamma^\mu C \) type, the mass ranges from \( 5.56 - 5.76 \, \text{GeV} \); for the \( C\gamma_\mu \gamma_5 \otimes \gamma_5 \gamma^\mu C \) type, the mass is between \( 5.38 - 5.62 \, \text{GeV} \); and for the \( C \otimes C \) type, the mass is in the range of \( 5.62 - 5.84 \, \text{GeV} \). We find that the masses of the \( C\gamma_5 \otimes \gamma_5 C \) and \( C\gamma_\mu \otimes \gamma^\mu C \) types are almost identical, while the \( C\gamma_\mu \gamma_5 \otimes \gamma_5 \gamma^\mu C \) type has the lowest mass, and the \( C \otimes C \) type has the highest mass. This could assist in differentiating them in upcoming high-energy nuclear and particle experiments.

\begin{acknowledgments}
We thank Zhi-Gang Wang for helpful discussions. This work is supported by the Natural Science Foundation of Hebei Province under Grant No. A2023205038.
\end{acknowledgments}

\begin{widetext}
\appendix
\appendix
\section{The spectral density \texorpdfstring{$\rho_{\text{OPE}}(s)$}{rho\_OPE(s)}}
\label{Appendix_A:_Spectral_densities_}
In this appendix, as examples, we present the OPE spectral densities \(\rho_{1}\) and \(\rho_{4}\) respectively derived from currents \(j_{1}\) and \(j_{4}\), considering only the triply charm tetraquark state \(cc\bar{c}\bar{s}\) . The spectral density of the triply charm tetraquark state \(cc\bar{c}\bar{q}\) can be obtained by making a simple replacement in the spectral density of the state \(cc\bar{c}\bar{s}\), where we set \(m_s \rightarrow 0\), \(\langle \bar{s}s \rangle \rightarrow \langle \bar{q}q \rangle\), and \(\langle \bar{s}g_s \sigma G s \rangle \rightarrow \langle \bar{q}g_s \sigma G q \rangle\). We define: \(G_{xy}\equiv m_c^2 \left( \frac{1}{x} + \frac{1}{y} + \frac{1}{1-x-y} \right) - s\), \(E_{xyz}\equiv m_c^2 \left( \frac{1}{x} + \frac{1}{y} + \frac{1}{z} \right) - s\), \(\tilde{m}_c^2\equiv m_c^2 \left( \frac{1}{x} + \frac{1}{y} + \frac{1}{1-x-y} \right)\), \(\overline{m}_c^2\equiv m_c^2 \left( \frac{1}{x} + \frac{1}{y} + \frac{1}{z} \right)\).\,\,The integration limits are defined as \(x_{\text{min}} \equiv \frac{-3 m_c^2 + s - \sqrt{9 m_c^4 - 10 m_c^2 s + s^2}}{2s}\), \(x_{\text{max}} \equiv \frac{-3 m_c^2 + s + \sqrt{9 m_c^4 - 10 m_c^2 s + s^2}}{2s}\), \(y_{\text{min}} \equiv \frac{1 - x - \sqrt{4 (m_c^2 x - m_c^2 x^2)/ (m_c^2 - s x) + (1 - x)^2}}{2}\), \(y_{\text{max}} \equiv \frac{1 - x + \sqrt{4 (m_c^2 x - m_c^2 x^2)/ (m_c^2 - s x) + (1 - x)^2}}{2}\), and \(z_{\text{min}} \equiv \frac{m_c^2 x y}{s x y - m_c^2 (x + y)}\).
It is noteworthy that when \(\delta(s - \tilde{m}_c^2)\) and \(\delta(s - \overline{m}_c^2)\) appear, we have \(\int_{x_{min}}^{x_{max}} dx \int_{y_{min}}^{y_{max}} dy \int_{z_{min}}^{1-x-y} dz \rightarrow \int_{0}^{1} dx \int_{0}^{1-x} dy \int_{0}^{1-x-y} dz\), and \(\int_{x_{min}}^{x_{max}} dx \int_{y_{min}}^{y_{max}} dy \rightarrow \int_{0}^{1} dx \int_{0}^{1-x} dy\). The OPE spectral density \(\rho_{1}(s)\) calculated from the current \(j_{1}\) is:

\begin{equation}
    \rho_{1}(s) = \rho^{(0)}_{1}(s) + \rho^{(3)}_{1}(s) + \rho^{(4)}_{1}(s) + \rho^{(5)}_{1}(s) + \rho^{(7)}_{1}(s) + \rho^{(8)}_{1}(s) + \rho^{(9)}_{1}(s) \, , \tag{A.1}
\end{equation}
where the superscript $(n)$ represents the dimension of the vacuum condensate term, where $n$ takes the values $0$, $3$, $4$, $5$, $7$, $8$, $9$. In our calculation, we found that the contribution of the three-gluon condensate $\langle g_s^3GGG \rangle$ is not only very small but also has a lengthy expression. Therefore, we do not present the analytical results of this part here. Hence, the explicit expression for the spectral density $\rho_1(s)$ is:

\begin{equation*}
\begin{aligned}
\rho^{(0)}_{1}(s)&=\int^{x_{max}}_{x_{min}}dx\int^{y_{max}}_{y_{min}}dy\int^{1-x-y}_{z_{min}}dz \frac{E_{xyz}^2 m_c^2 \left(3 z^2 m_c m_s + E_{xyz} x^3\right)}{64 \pi^6 z^2}\\
&+
   \int^{x_{max}}_{x_{min}}dx\int^{y_{max}}_{y_{min}}dy \frac{G_{xy}^3 x^4 \left(4 m_c m_s + \frac{G_{xy} x^3}{(x+y-1)^2}\right)}{256 \pi^6 y^2}\, ,
\end{aligned}
\end{equation*}

\begin{equation*}
\begin{aligned}
  \rho^{(3)}_{1}(s) & = \int_{x_{min}}^{x_{max}}dx\int_{y_{min}}^{y_{max}} dy \,
  \Bigg\{ \frac{ \langle \bar{s}s\rangle}{16 \pi^4 y^2 (x+y-1)^2} \Big\{ 2 G_{xy} x^4 m_c (G_{xy} - s) (x+y-1)^2   \\
  & \quad     +x^3 y^2 m_c^2 (2 G_{xy} - s) m_s  + 2 G_{xy} y^2 m_c^3 (x+y-1)^2 + x^7 (3 G_{xy}^2 - 6 G_{xy} s + s^2) m_s \Big\} \Bigg\},  
\end{aligned}
\end{equation*}

\begin{equation*}
\begin{aligned}
\rho^{(4)}_{1}(s)&= \int^{x_{max}}_{x_{min}}dx\int^{y_{max}}_{y_{min}}dy\int^{1-x-y}_{z_{min}}dz \Bigg\{\frac{\langle \frac{\alpha_s GG}{\pi}\rangle \,m_c^2}{384 \pi^4 x^2 y^3 z^5} \Big\{ 3 y z^3 m_c m_s \left(2 x^2 (y^2+z^2) - x y z^2 + 2 y^2 z^2\right)   \\
&  + 2 m_c^2 \left(x^5 (y^3+z^3) + x^2 y^3 z^3\right) + 3 E_{xyz} x^3 y z^3 \left(2 x^2 - x y + 2 y^2\right) \Big\} \\
& - \frac{\langle \frac{\alpha_s GG}{\pi}\rangle \, m_c^5 \, m_s \left(x^3 (y^3+z^3) + y^3 z^3\right) \delta \left(s - \overline{m}_c^2\right)}{192 \pi^4 x^3 y^3 z^3} \Bigg\} \\
&+
   \int^{x_{max}}_{x_{min}}dx\int^{y_{max}}_{y_{min}}dy \Bigg\{ \frac{\langle \frac{\alpha_s GG}{\pi}\rangle \, x}{1536 \pi^4 y^5 (x+y-1)^5} \Big\{ -6 G_{xy} x^2 y^2 m_c m_s (x+y-1)^3 (x^2 - 2x(y+1)\\
   &+ (y-1)^2) + 2 G_{xy} x^2 m_c^2 \left( 12(x-1)x^4 y^2 + 12(x-1)^2 x^4 y + 4(x-1)^3 x^4 + 2(5x-3) y^6 \right. \\
&+ 3(x-1)(5x-1) y^5 + 12(x-1)^2 x y^4 + 4(x-1)^3 x y^3 + 3 y^7 ) \\
&+ 8 m_c^3 m_s (x+y-1)^2 (x^6 + 3x^5(y-1) + 3x^4(y-1)^2 + x^3(y-1)^3 \\
&\left. + 3x^2(y-1)y^3 + 3x(y-1)^2 y^3 + (y-1)^3 y^3 \right) \\
& + 3 G_{xy} x^5 y^2 (x+y-1)^2 (G_{xy} x (2y-1) - y\, G_{xy}  + G_{xy} - 2sxy) \Big\} \Bigg\},
\end{aligned}
\end{equation*}
\begin{equation*}
\begin{aligned}
    \rho^{(5)}_{1}(s) &=\int^{x_{max}}_{x_{min}}dx\int^{y_{max}}_{y_{min}}dy \Bigg\{
    \frac{\langle \bar{s}g_s \sigma G s \rangle \tilde{m}_c^2 \delta(s-\tilde{m}_c^2)}{96 \pi^4  M_B^2 \, y^2 (x+y-1)^2}   \Big\{x^7 \tilde{m}_c^2 (-m_s) (\tilde{m}_c^2 + 9 M_B^2 \,) \\
    &+ x^3 y^2 m_c^2 m_s (\tilde{m}_c^2 + 4 M_B^2 \,) - 3 M_B^2 \, x^4 m_c \tilde{m}_c^2 (x+y-1)^2 + 3 M_B^2 \, y^2 m_c^3 (x+y-1)^2 \Big\} \\
&+ \frac{\langle \bar{s}g_s \sigma G s \rangle m_s \delta(s-\tilde{m}_c^2) (x^7 \tilde{m}_c^4 - x^3 y^2 m_c^2 \tilde{m}_c^2)}{128 \pi^4 y^2 (x+y-1)^3} \\
&+ \frac{\langle \bar{s}g_s \sigma G s \rangle}{64 \pi^4 y^2 (x+y-1)^3} \Big\{ x^4 m_c (x+y-1)^2 (s (1 - 12  (x+y-1)) - 2 G_{xy} (1 - 6  (x+y-1)))  \\
& + x^3 y^2 m_c^2 m_s (4  (x+y-1) - 1) - y^2 m_c^3 (x+y-1)^2 (1 - 4  (x+y-1)) \\ 
&+ x^7 m_s (G_{xy} (16  (x+y-1) - 3 ) + 3s (1 - 8  (x+y-1))) \Big\}  \Bigg\}
,
\end{aligned}
\end{equation*}
\begin{equation*}
\begin{aligned}
    \rho^{(7)}_{1}(s) & =\int^{x_{max}}_{x_{min}}dx\int^{y_{max}}_{y_{min}}dy \Bigg\{ -\left(\frac{\langle \frac{\alpha_s GG}{\pi}\rangle \, \langle \bar{s}s \rangle \left(3 m_s x^3 + 2 \left(x^2 - 2(y+1)x + (y-1)^2\right) m_c\right) x^3}{192 \pi^2 y^3 (x+y-1)^2}\right) \\
& - \frac{\langle \frac{\alpha_s GG}{\pi}\rangle \, \langle \bar{s}s\rangle\, \delta\left(s-\tilde{m}_c^2\right)}{1152 (M_B^2)^2 \, \pi^2 y^5 (x+y-1)^5 x^3}  \bigg\{ 3 M_B^2 \, y^2 (x+y-1)^3 m_s \tilde{m}_c^2 \left(\tilde{m}_c^2 + 4 M_B^2 \,\right) x^9  \\
& + 6 (M_B^2)^2 y^2 (x+y-1)^3 \left(x^2 - 2(y+1)x + (y-1)^2\right) m_c \tilde{m}_c^2 x^6 \\
& + 2 m_c^2 m_s (2 x^3 (y^6 + 3(x-1)y^5 + 3(x-1)^2 y^4 + (x-1)^3 y^3 + 3(x-1) x^3 y^2 + 3(x-1)^2 x^3 y \\
&+ (x-1)^3 x^3) \tilde{m}_c^4  + M_B^2 \, \left(6 y^8 + 3(5x-6) y^7 + (x(x(4x+15)-27)+18) y^6 \right.\\
&+ 3(x-1)(x(x(4x+5)-1)+2) y^5 + 3(x-1)^2 x(x+1)(4x+1) y^4 + 2(x-1)^3 x^2(2x+3) y^3 \\
&+ 12(x-1) x^6 y^2 + 12(x-1)^2 x^6 y \\ 
&+ 4(x-1)^3 x^6   \left. \right) \tilde{m}_c^2 + M_B^4(6y^8 +3(5x-6)y^7 +(x(x(4x+15)-27)+18)y^6 \\
&+3(x-1)(x(x(4x+5)-1)+2)y^5 +3(x-1)^2 x(x+1)(4x+1)y^4+2(x-1)^3 x^2 (2x+3)y^3 \\
&+12(x-1)x^6 y^2+12(x-1)^2 x^6 y+4(x-1)^3 x^6 ) )  x^4  \\
& - 4 y^2 (x^6 + 3(y-1) x^5 + 3(y-1)^2 x^4 + (y-1)^3 x^3 + 3(y-1) y^3 x^2 + 3(y-1)^2 y^3 x \\
&+ (y-1)^3 y^3) m_c^4 m_s \tilde{m}_c^2 x^3 \\
& + 4 M_B^2 \, (x+y-1)^2 m_c^3 \, x \Big\{  2 \left(y^6 + 3(x-1) y^5 + 3(x-1)^2 y^4 + (x-1)^3 y^3 + 3(x-1) x^3 y^2 \right.\\
&+ 3(x-1)^2 x^3 y \left. + (x-1)^3 x^3\right) \tilde{m}_c^2 x^3  \\
& + M_B^2 \, \left(6 y^8 + 3(5x-6) y^7 + (x(x(2x+21)-27)+18) y^6 + 3(x-1)(x(x(2x+7)-1)+2) y^5  \right.\\
& + 3(x-1)^2 x(x(2x+5)+1) y^4 + 2(x-1)^3 x^2(x+3) y^3 + 6(x-1) x^6 y^2 + 6(x-1)^2 x^6 y \\
&+ 2(x-1)^3 x^6  \left. \right)  \Big\}  - 8 M_B^2 \, y^2 (x+y-1)^2 \left(x^6 + 3(y-1) x^5 + 3(y-1)^2 x^4 + (y-1)^3 x^3 \right.\\
&+ 3(y-1) y^3 x^2 + 3(y-1)^2 y^3 x  \left.+ (y-1)^3 y^3\right) m_c^5
\bigg\} \Bigg\},
\end{aligned}
\end{equation*}
\begin{equation*}
\begin{aligned}
    \rho^{(8)}_{1}(s)&=\int^{x_{max}}_{x_{min}}dx\int^{y_{max}}_{y_{min}}dy\int^{1-x-y}_{z_{min}}dz \Bigg\{  \frac{ \langle \frac{\alpha_s GG}{\pi}\rangle ^2\, m_c^2 \delta\left(s-\overline{m}_c^2\right)}{6912 \pi^2 (M_B^2)^3 x^3 y^3 z^5} \bigg\{ z^2 m_c m_s \\
    &\times \left(-3 M_B^2 \,  m_c^2 \left(2 z (x^3+y^3) + x y (2 x^2 - x y + 2 y^2) + 2 z^3 (x+y)\right) + 2 m_c^4 (x^3+y^3+z^3) \right. \\ &\left.
    + 9 (M_B^2)^2 \, x y z (2 x^2 - x y + 2 (y^2 + z^2))\right)  \\
& + M_B^2 \, x^3 \left(3 M_B^2 \, m_c^2 (x y (2 x^2 - x y + 2 y^2) + 2 z^3 (x+y)) - 2 m_c^4 (x^3+y^3+z^3) - 18 (M_B^2 )^2 \,  x y z^3\right) \bigg\}
 \Bigg\} \\
 &+\int^{x_{max}}_{x_{min}}dx\int^{y_{max}}_{y_{min}}dy \Bigg\{ -\frac{\langle \frac{\alpha_s GG}{\pi}\rangle ^2\, \delta\left(s-\tilde{m}_c^2\right)}{55296 \pi^2 (M_B^2)^2 \, y^5 (x+y-1)^5}  \bigg\{-36 (M_B^2)^2 \, x^3 y^2 m_c m_s  \\
 & \times (x+y-1)^3 + 9 (M_B^2)^2 \, x^6 y^2 \tilde{m}_c^2 (x+y-1)^2 \left.- 12 M_B^2 \,  x m_c^3 m_s (x+y-1)^2 (4 x^4 + 4 x^3 (y-1) \right.\\
 &+ x^2 y^2 + 4 x y^3 + 4 (y-1) y^3)  \\
&\left.+ 6 M_B^2 \, x m_c^2 (2 x^3 \tilde{m}_c^2 (x+y-1)^2 (x^3+y^3) + M_B^2 \, (4 (x-1) x^6 y + 2 (x-1)^2 x^6 + 2 x^5 (x+1) y^2 \right. \\
&\left.+ 2 (x-1)^2 x^2 (x+3) y^3 + 3 (3x-4) y^6 + 2 \left(x(x(x+3)-3)+3\right) y^5 + (x-1) x (x(4x+9)+3) y^4 \right.\\
&+ 6 y^7) )  \left.- 4 M_B^2 \, m_c^4 \left(12 x^6 (y-1) + 3 x^5 (y(5y-8)+4) + 2 x^4 \left(3(y-2)(y-1)y-2\right) + 3 x^3 (y-1)^2 y^2 \right.\right. \\
&\left.\left.+ 3 x^2 y^5 + 6 x (y-1) y^5 + 3 (y-1)^2 y^5\right) \right. \\
&+ 16 x m_c^5 m_s (x+y-1)^2 \left(3 x^2 (y-1) + 3 x (y-1)^2 - 3 (y-1) y - 1\right)\bigg\} - \frac{\langle \frac{\alpha_s GG}{\pi}\rangle ^2\, x^6}{3072 \pi^2 y^3 (x+y-1)^3}
 \Bigg\},
\end{aligned}
\end{equation*}
\begin{equation*}
\begin{aligned}
    \rho^{(9)}_{1}(s)&=\int^{x_{max}}_{x_{min}}dx\int^{y_{max}}_{y_{min}}dy \Bigg\{ \frac{\langle \frac{\alpha_s GG}{\pi}\rangle \langle \bar{s}g_s \sigma G s \rangle \delta\left(s-\tilde{m}_c^2\right)}{6912 \pi^2 (M_B^2)^4 \, x^3 y^5 (x+y-1)^5}   \bigg\{ 3 M_B^2 \, y^2 (x+y-1)^3  \\ 
    & m_s  \left(\tilde{m}_c^6 + 3 M_B^2 \, \tilde{m}_c^4 + 6 (M_B^2)^2 \, \tilde{m}_c^2 + 6 (M_B^2)^3 \, \right) x^9  \\
&+ 9 (M_B^2)^2\,  y^2 (x+y-1)^3 \left(x^2 - 2(y+1) x + (y-1)^2\right) m_c \left(\tilde{m}_c^4 + 2 M_B^2 \, \tilde{m}_c^2 + 2 (M_B^2)^2\,\right) x^6 \\
&+ 2 m_c^2 m_s \tilde{m}_c^4 \left( 2 (y^6 + 3(x-1) y^5 + 3(x-1)^2 y^4 + (x-1)^3 y^3 + 3(x-1) x^3 y^2 + 3(x-1)^2 x^3 y \right.\\
&+ (x-1)^3 x^3) \tilde{m}_c^2 x^3   \left. + 3 M_B^2 \, y^3 (x+y-1)^3 (2 x^2 - y x + 2 y^2) \right) x^4 \\
&- 4 y^2 \left(x^6 + 3 (y-1) x^5 + 3 (y-1)^2 x^4 + (y-1)^3 x^3 + 3 (y-1) y^3 x^2 + 3 (y-1)^2 y^3 x + (y-1)^3 y^3\right) \\ 
&\times m_c^4 m_s \left(\tilde{m}_c^2 - 2 M_B^2 \,\right) x^3 \\
&+ 6 M_B^2 \, (x+y-1)^2 m_c^3 \tilde{m}_c^2 \left( 2 (x^6 + 3 (y-1) x^5 + 3 (y-1)^2 x^4 + (y-1)^3 x^3 + 3 (y-1) y^3 x^2 \right. \\
&+ 3 (y-1)^2 y^3 x + (y-1)^3 y^3) \tilde{m}_c^2 x^3  \\
&\left. \left. + 3 M_B^2 \, y^3 (x+y-1) (2 x^4 + (3 y-4) x^3 + (4 y^2 - 2 y + 2) x^2 + (y-1) y (3 y + 1) x + 2 (y-1)^2 y^2) \right) x \right. \\
&+ 12 M_B^2 \, y^2 (x+y-1)^2 \left(x^6  + 3 (y-1) x^5 + 3 (y-1)^2 x^4 + (y-1)^3 x^3 + 3 (y-1) y^3 x^2  \right. \\
&\left. + 3 (y-1)^2 y^3 x + (y-1)^3 y^3\right) m_c^5 \left(M_B^2 \, - \tilde{m}_c^2\right)  \bigg\} \\
&- \frac{\langle \frac{\alpha_s GG}{\pi}\rangle \langle \bar{s}g_s \sigma G s \rangle \delta\left(s-\tilde{m}_c^2\right)}{9216 (M_B^2)^3\, \pi^2 x^3 y^5 (x+y-1)^3}  \bigg\{3 M_B^2 \, y^2 m_s \\
&\times \left(\tilde{m}_c^4 + 2 M_B^2 \, \tilde{m}_c^2 + 2 (M_B^2)^2\,\right) x^9 + 6 (M_B^2)^2 \, y^2 (x+y-1)^2 m_c \left(\tilde{m}_c^2 + M_B^2 \,\right) x^6 \\
&+ 2 m_c^2 m_s \tilde{m}_c^2 \left(2 (x^3+y^3) \tilde{m}_c^2 x^3 + 3 M_B^2 \, y^3 (2 x^2 - y x + 2 y^2) \right) x^4 + 4 y^2 (x^3+y^3) m_c^4 m_s \left(M_B^2 \, - \tilde{m}_c^2\right) x^3 \\
&+ 4 M_B^2 \, (x+y-1)^2 m_c^3 \left(2 (x^3+y^3) \tilde{m}_c^2 x^3 + 3 M_B^2 \, y^3 (2 x^2 - y x + 2 y^2) \right) x \\
&- 8 M_B^2 \, y^2 (x+y-1)^2 (x^3+y^3) m_c^5 \bigg\}
  \Bigg\}.
\end{aligned}
\end{equation*}

The spectral density \(\rho_{4}(s)\) derived from the current \(j_{4}\) is:

\begin{equation}
    \rho_{4}(s) = \rho^{(0)}_{4}(s) + \rho^{(3)}_{4}(s) + \rho^{(4)}_{4}(s) + \rho^{(5)}_{4}(s) + \rho^{(7)}_{4}(s) + \rho^{(8)}_{4}(s) + \rho^{(9)}_{4}(s) \, , \tag{A.2}
\end{equation}
where the superscript $(n)$ represents the dimension of the vacuum condensate term, where $n$ takes the values $0$, $3$, $4$, $5$, $7$, $8$, $9$. The specific spectral density is:

\begin{equation*}
\begin{aligned}
    \rho^{(0)}_{4}(s)&=\int^{x_{max}}_{x_{min}}dx\int^{y_{max}}_{y_{min}}dy\int^{1-x-y}_{z_{min}}dz \Bigg\{ -\frac{E_{xyz}^2 m_c^2 \left(E_{xyz} x^3 - 6 z^2 m_c m_s\right)}{64 \pi^6 z^2}
 \Bigg\} \\
&+
   \int^{x_{max}}_{x_{min}}dx\int^{y_{max}}_{y_{min}}dy \,\Bigg\{ \frac{G_{xy}^3 x^4 \left(G_{xy} x^3 - 2 m_c m_s (x+y-1)^2\right)}{128 \pi^6 y^2 (x+y-1)^2}
 \Bigg\} ,
\end{aligned}
\end{equation*}
\begin{equation*}
\begin{aligned}
    \rho^{(3)}_{4}(s)&=\int^{x_{max}}_{x_{min}}dx\int^{y_{max}}_{y_{min}}dy \,\Bigg\{ \frac{\langle \bar{s}s\rangle }{16 \pi^4 y^2 (x+y-1)^2}\bigg\{-2 G_{xy} x^4 m_c (G_{xy}-s) (x+y-1)^2 \\
    &+ x^3 y^2 m_c^2 (s-2 G_{xy}) m_s 
    + 4 G_{xy} y^2 m_c^3 (x+y-1)^2 + 2 x^7 (3 G_{xy}^2 - 6 G_{xy} s + s^2) m_s\bigg\}
\Bigg\} ,
\end{aligned}
\end{equation*}
\begin{equation*}
\begin{aligned}
\rho^{(4)}_{4}(s)&=\int^{x_{max}}_{x_{min}}dx\int^{y_{max}}_{y_{min}}dy\int^{1-x-y}_{z_{min}}dz \Bigg\{ \frac{\langle \frac{\alpha_s GG}{\pi}\rangle \, m_c^2 }{768 \pi^4 x^2 y^3 z^5}\bigg\{x^2 \left(-4 m_c^2 \left(x^3 (y^3 + z^3) + y^3 z^3\right) \right.\\
&\left.- 3 E_{xyz} x y z^2 \left(4 z (x^2 + y^2) +  x y (x + y)\right)\right) + 6 y z^3 m_c m_s \left(4 z^2 (x^2 + y^2) + 4 x^2 y^2 +  x y z (x + y)\right)\bigg\}\\
&- \frac{\langle \frac{\alpha_s GG}{\pi}\rangle \, m_c^5 \, m_s \left(x^3 (y^3 + z^3) + y^3 z^3\right) \delta\left(s - \overline{m}_{c}^2\right)}{96 \pi^4 x^3 y^3 z^3}
\Bigg\} \\
&+
   \int^{x_{max}}_{x_{min}}dx\int^{y_{max}}_{y_{min}}dy \,\Bigg\{\frac{\langle \frac{\alpha_s GG}{\pi}\rangle \, x}{1536 \pi^4 y^5 (x+y-1)^5}  \bigg\{6 G_{xy}\,x^2 y^2 m_c m_s (x + y - 1)^3 (2 x y + x (3 x - 5) \\
   &+ 3 y^2 - 5 y + 2) + 
2 G_{xy} x m_c^2 (24 (x - 1) x^5 y^2 + 24 (x - 1)^2 x^5 y + 8 (x - 1)^3 x^5 + 
8 (x - 1)^3 x^2 y^3\\
&+ 9 (1 - 2 x) y^7 + ((39 - 22 x) x - 9) y^6 + 3 (x - 1) (x (2 x + 7) - 1) y^5 + 
3 (x - 1)^2 x (7 x + 1) y^4 - 3 y^8)\\
&- 
8 m_c^3 m_s (x + y - 1)^2 (x^6 + 3 x^5 (y - 1) + 3 x^4 (y - 1)^2 + 
x^3 (y - 1)^3 + 3 x^2 (y - 1) y^3 + 3 x (y - 1)^2 y^3 \\
&+ (y - 1)^3 y^3) + 
3 G_{xy} x^5 y^2 (x + y - 1)^2 (x + y) (G_{xy} (2 x + 2 y - 3) - 2 s (x + y - 1))
\bigg\}
 \Bigg\},
\end{aligned}
\end{equation*}
\begin{equation*}
\begin{aligned}
\rho^{(5)}_{4}(s)&=\int^{x_{max}}_{x_{min}}dx\int^{y_{max}}_{y_{min}}dy \,\Bigg\{\frac{\langle \bar{s}g_s \sigma G s \rangle \tilde{m}_c^2 \delta\left(s - \tilde{m}_c^2\right) }{96 \pi^4 M_B^2 \, y^2 (x+y-1)^2}\bigg\{-2 x^7 \tilde{m}_c^2 m_s \left(\tilde{m}_c^2 + 9 \, M_B^2 \,\right) \\
&- x^3 y^2 m_c^2 m_s \left(\tilde{m}_c^2 + 4 \, M_B^2 \,\right) + 3 \, M_B^2 \, x^4 m_c \tilde{m}_c^2 (x+y-1)^2 + 6 \, M_B^2 \, y^2 m_c^3 (x+y-1)^2 \bigg\}\\
&+ \frac{\langle \bar{s}g_s \sigma G s \rangle}{16 \pi^4 y^2 (x+y-1)^2
}\bigg\{x^3 m_s (4 x^4 (2 G_{xy} - 3 s) - y^2 m_c^2) - m_c (x + y - 1)^2 (3 x^4 (G_{xy} - s) - 2 y^2 m_c^2)
\bigg\}
 \Bigg\} ,
\end{aligned}
\end{equation*}
\begin{equation*}
\begin{aligned}
\rho^{(7)}_{4}(s)&=\int^{x_{max}}_{x_{min}}dx\int^{y_{max}}_{y_{min}}dy \,\Bigg\{\frac{\langle \frac{\alpha_s GG}{\pi}\rangle \, \langle \bar{s}s\rangle \, x^3}{192 \pi^2 y^3 (x+y-1)^3}\bigg\{2 (x + y - 1) (3 y^2 + 2 x y - 5 y + x (3 x - 5) + 2) m_c \\
&- 3 x^3 (x + y) m_s
\bigg\}  \\
&- \frac{\langle \frac{\alpha_s GG}{\pi}\rangle \, \langle \bar{s}s\rangle \, \delta\left(s - \tilde{m}_c^2\right)}{1152 (M_B^2 )^2 \,  \pi^2 y^5 (x+y-1)^5 x^3}  \bigg\{3 y^2 (x + y - 1)^2 (x + y) m_s M_B^2 \tilde{m}_c^2 (4 M_B^2 + \tilde{m}_c^2) x^9 \\
&- 6 y^2 (x + y - 1)^3 (3 y^2 + 2 x y - 5 y + x (3 x - 5) + 2) m_c M_B^4 \tilde{m}_c^2 x^6 + 
m_c^2 m_s ((-12 y^8 + 3 (12 - 11 x) y^7 \\
&+ (x ((x ((16 x - 39) + 66)) - 36)) y^6 \\
&+ 3 (x - 1) (x (x (16 x - 13) + 7) - 4) y^5 + 
3 (x - 1)^2 x^2 (16 x - 11) y^4 + 4 (x - 1)^3 x^2 (4 x - 3) y^3 \\
&+ 48 (x - 1) x^6 y^2 + 48 (x - 1)^2 x^6 y + 16 (x - 1)^3 x^6) M_B^4 + 
(-12 y^8 + 3 (12 - 11 x) y^7 \\
&+ (x (x (16 x - 39) + 66) - 36) y^6 \\
&+ 3 (x - 1) (x (x (16 x - 13) + 7) - 4) y^5 + 
3 (x - 1)^2 x^2 (16 x - 11) y^4 + 4 (x - 1)^3 x^2 (4 x - 3) y^3 \\
&+ 48 (x - 1) x^6 y^2 + 48 (x - 1)^2 x^6 y + 16 (x - 1)^3 x^6) 
\tilde{m}_c^2 M_B^2 + 8 x^3 (y^6 + 3 (x - 1) y^5 + 3 (x - 1)^2 y^4\\
&+ (x - 1)^3 y^3 + 3 (x - 1) x^3 y^2 + 3 (x - 1)^2 x^3 y + (x - 1)^3 x^3) 
\tilde{m}_c^4)x^4 + 4y^2(x^6+3(y-1)x^5\\
&+3(y-1)^2 x^4+(y-1)^3 x^3 + 3(y-1)y^3x^2+3(y-1)^2y^3x+(y-1)^3 y^3)m_c^4 m_s \tilde{m}_c^2 x^3 \\
&- 4 (x + y - 1)^2 m_c^3 M_B^2 (2 (y^6 + 3 (x - 1) y^5 + 3 (x - 1)^2 y^4 + (x - 1)^3 y^3 + 3 (x - 1) x^3 y^2 \\
&+ 
3 (x - 1)^2 x^3 y + (x - 1)^3 x^3) \tilde{m}_c^2 x^3 + (-12 y^8 + 3 (12 - 11 x) y^7 + (x (x (2 x - 51) + 66) - 36) y^6 \\
&+ 
3 (x - 1) (x (x (2 x - 17) + 7) - 4) y^5 + 3 (x - 1)^2 x^2 (2 x - 11) y^4 + 2 (x - 6) (x - 1)^3 x^2 y^3 \\
& + 
6 (x - 1) x^6 y^2 + 6 (x - 1)^2 x^6 y + 2 (x - 1)^3 x^6) M_B^2) x \\
&- 16 y^2 (x + y - 1)^2 (x^6 + 3 (y - 1) x^5 + 
3 (y - 1)^2 x^4 + (y - 1)^3 x^3 + 3 (y - 1) y^3 x^2 + 3 (y - 1)^2 y^3 x \\
&+ (y - 1)^3 y^3) m_c^5 M_B^2\bigg\}
 \Bigg\} ,
\end{aligned}
\end{equation*}
\begin{equation*}
\begin{aligned}
\rho^{(8)}_{4}(s)&=\int^{x_{max}}_{x_{min}}dx\int^{y_{max}}_{y_{min}}dy\int^{1-x-y}_{z_{min}}dz \Bigg\{\frac{\langle \frac{\alpha_s GG}{\pi}\rangle ^2 m_c^2 \delta\left(s - \overline{m}_c^2\right)}{13824 \pi^2 (M_B^2)^3\,  x^3 y^3 z^5}  \bigg\{9 x^4 y z^2 (x + y + 4 z) M_B^6 \\
&- 
3 x^3 (4 (x + y) z^3 + (x^2 + y^2) z^2 + 4 x y (x^2 + y^2)) m_c^2 M_B^4 + 
18 x y z^3 (4 z^2 + (x + y) z \\
&+ 4 (x^2 + y^2)) m_c m_s M_B^4 + 
4 x^3 (x^3 + y^3 + z^3) m_c^4 M_B^2 - 
6 z^2 (4 (x + y) z^3 + (x^2 + y^2) z^2 + 4 (x^3 + y^3) z \\
&+ 4 x y (x^2 + y^2)) 
m_c^3 m_s M_B^2 + 
8 z^2 (x^3 + y^3 + z^3) m_c^5 m_s
\bigg\}
 \Bigg\} \\
&+
   \int^{x_{max}}_{x_{min}}dx\int^{y_{max}}_{y_{min}}dy \,\Bigg\{ -\frac{ \langle \frac{\alpha_s GG}{\pi}\rangle^2 \, \delta\left(s - \tilde{m}_c^2\right)}{13824 \pi^2 (M_B^2)^2\, y^5 (x+y-1)^5} \bigg\{6 y^2 (x + y - 1)^2 M_B^4 \tilde{m}_c^2 x^6 \\
   &+ 
18 y^2 (x + y - 1)^3 m_c m_s M_B^4 x^3 + 
3 (x + y - 1)^2 (3 x^4 + 2 (y - 1) x^3 + (2 y - 1) x^2 + 2 y^2 (y + 1) x \\
&+ (y - 1) y^2 (3 y + 1)) 
m_c^3 m_s M_B^2 x - 
4 (x + y - 1)^2 (3 (y - 1) x^2 + 3 (y - 1)^2 x \\
&- 3 (y - 1) y - 1) 
m_c^5 m_s x + 
3 m_c^2 M_B^2 ((y^5 + 3 (x - 1) y^4 + 3 (x - 1)^2 y^3 + (3 (x - 1)^2 x - 1) y^2 \\
&+ 3 (x - 1)^2 x^2 y + (x - 1)^3 x^2) 
\tilde{m}_c^2 x^3 + (-9 y^7 + 3 (7 - 8 x) y^6 + (x ((x - 30) x + 39) - 15) y^5\\
&+ 
(x (x (x (3 x - 28) + 39) - 18) + 3) y^4 + 
(x - 1) x (x (x (3 x - 14) + 12) - 3) y^3 \\
&+ 
(x - 2) x^3 (x (3 x - 2) + 1) y^2 + 
(x - 1) x^5 (3 x - 5) y + (x - 2) (x - 1)^2 x^5) 
M_B^2) x \\
&+ 
(9 y^7 + 21 (x - 1) y^6 + 15 (x - 1)^2 y^5 + 3 (x (x (5 x - 6) + 3) - 1) y^4 + 
3 (x - 1) x^2 (7 x - 3) y^3 \\
&- 3 (x - 1) x^2 (x + 1) (5 x - 1) y^2 - 
24 (x - 1)^2 x^4 y + 8 x^4 (3 (x - 1) x + 1)) 
m_c^4 M_B^2 \bigg\} \\
&- \frac{\langle \frac{\alpha_s GG}{\pi}\rangle ^2 \, x^6}{1152 \pi^2 y^3 (x+y-1)^3}
\Bigg\},
\end{aligned}
\end{equation*}
\begin{equation*}
\begin{aligned}
\rho^{(9)}_{4}(s)&=\int^{x_{max}}_{x_{min}}dx\int^{y_{max}}_{y_{mi}}dy \,\Bigg\{\frac{\langle \frac{\alpha_s GG}{\pi}\rangle \, \langle \bar{s}g_s \sigma G s \rangle \, \delta\left(s - \tilde{m}_c^2\right)}{6912 \pi^2 (M_B^2)^4\, x^3 y^5 (x+y-1)^5}  \\
& \times \bigg\{3 x^9 y^2 M_B^2 m_s (x + y - 1)^2 (x + y) (6 M_B^4 \tilde{m}_c^2 + 3 M_B^2 \tilde{m}_c^4 + 6 M_B^6 + \tilde{m}_c^6) \\
&+ 
x^4 m_c^2 \tilde{m}_c^4 m_s (8 x^3 \tilde{m}_c^2 (3 (x - 1) x^3 y^2 + 3 (x - 1)^2 x^3 y + (x - 1)^3 x^3 + 3 (x - 1) y^5 + 3 (x - 1)^2 y^4 \\
&+ (x - 1)^3 y^3 + y^6) - 
3 y^3 M_B^2 (x + y - 1)^2 (3 x^2 y + 4 (x - 1) x^2 + (3 x - 4) y^2 + 4 y^3)) \\
&+ 
4 x^3 y^2 m_c^4 \tilde{m}_c^2 m_s (x^6 + 3 x^5 (y - 1) + 3 x^4 (y - 1)^2 + x^3 (y - 1)^3 + 3 x^2 (y - 1) y^3 + 3 x (y - 1)^2 y^3 \\
&+ (y - 1)^3 y^3) ( 
\tilde{m}_c^2 - 2 M_B^2 ) - 
9 x^6 y^2 M_B^4 m_c (x + y - 1)^3 (2 x y + x (3 x - 5) + 3 y^2 - 5 y + 2)\\
&\times (2 M_B^2 \tilde{m}_c^2 + 2 M_B^4 + \tilde{m}_c^4) - 
6 x M_B^2 m_c^3 \tilde{m}_c^2 (x + y - 1)^2 (2 x^3 \tilde{m}_c^2 (3 (x - 1) x^3 y^2 + 3 (x - 1)^2 x^3 y \\
&+ (x - 1)^3 x^3 + 3 (x - 1) y^5 + 3 (x - 1)^2 y^4 + (x - 1)^3 y^3 + y^6) - 
3 y^3 M_B^2 (x + y - 1) (7 (x - 1) x^2 y \\
&+ 4 (x - 1)^2 x^2 + (7 x - 8) y^3 + (x (10 x - 7) + 4) y^2 + 4 y^4)) + 
24 y^2 M_B^2 m_c^5 (x + y - 1)^2 \\
& \times (x^6 + 3 x^5 (y - 1) + 3 x^4 (y - 1)^2 + x^3 (y - 1)^3 + 3 x^2 (y - 1) y^3 + 3 x (y - 1)^2 y^3 \\
&+ (y - 1)^3 y^3) ( 
M_B^2 - \tilde{m}_c^2 )
\bigg\}
\Bigg\}.
\end{aligned}
\end{equation*}
\end{widetext}

\nocite{*}
\bibliography{ref}

\end{document}